**Title:** Liquid solution centrifugation for safe, scalable, and efficient isotope separation

**Authors:** Joseph F. Wild[1*], Heng Chen[2*], Keyue Liang[1], Jiayu Liu[1], Stephen E. Cox[2], Alex N. Halliday[2†], Yuan Yang[1†]

**Affiliations:**

[1]Department of Applied Physics and Applied Mathematics, Columbia University, New York, NY 10027, United States

[2]Lamont-Doherty Earth Observatory, Columbia University, Palisades, NY 10964, United States

*These authors contributed equally to this work

[†]Corresponding author. Email: yy2664@columbia.edu, alexhalliday@climate.columbia.edu

**Abstract:** A general method of separating isotopes by centrifuging dissolved chemical compounds in a liquid solution is introduced. This technique can be applied to almost all elements and leads to large separation factors. The method has been demonstrated in several isotopic systems including Ca, Mo, O, and Li with single-stage selectivities of 1.046-1.067 per unit mass difference (e.g., 1.434 in $^{40}Ca/^{48}Ca$, 1.134 in $^{16}O/^{18}O$), which are beyond the capabilities of various conventional methods of isotope enrichment. Equations are derived to model the process and the results agree with those of the experiments. The scalability of the technique has been demonstrated by performing a three-stage enrichment of $^{48}Ca$, and the scalability is more broadly supported through analogies to the gas centrifuge, whereby countercurrent centrifugation can further multiply the separation factor by 5-10 times per stage in a continuous process. Optimal centrifuge conditions and solutions can achieve both high-throughput and highly efficient isotope separation.

**One-Sentence Summary:** Centrifuging a liquid solution can efficiently separate element's isotopes in both the solvent and the dissolved solutes.

**Main Text:**

The discovery of isotopes in the early 20th century led to countless world-changing technologies and applications. Enriched stable isotopes remain essential to help solve many of the most challenging questions in sustainability and fundamental science, such as $^{6}Li$ as the source for generating $^{3}H$ in nuclear fusion, $^{48}Ca$ as a key source for producing superheavy elements and examining the Standard Model, and $^{100}Mo$ as a precursor for $^{99m}Tc$ within the broad field of radiopharmaceuticals (*1*).

Various methods have been developed for efficient isotope enrichment, including gas centrifuge, electromagnetism, gas diffusion, chemical exchange, and laser separation (*2-7*). Each method has its own advantages and disadvantages. For example, gas centrifuges are very successful at separating isotopes that can form gaseous molecules at near-ambient temperatures, such as $UF_6$ for $^{235}U/^{238}U$, and $Ni(PF_3)_4$ for $^{62}Ni/^{64}Ni$ (*8*), but they are not suitable for isotopes that cannot be gasified at these temperatures, such as Group I and II elements. Moreover, most gaseous precursors are highly toxic. Electromagnetic isotope separation (EMIS) has near-perfect selectivity, but the production rate is extremely low and the cost prohibitively high, making it only suitable for isotopes with a mg - g / year demand. Chemical methods utilize the isotope effects in the Gibbs free energy (e.g., molecular and atomic vibrations), which has been successful for light isotopes particularly with a large ($1/M_1$-$1/M_2$) such as H/D and $^{6}Li/^{7}Li$. However, hazardous chemicals are often involved, such as $H_2S$ for H/D and Hg for $^{6}Li/^{7}Li$, and the effect weakens substantially for heavier elements (*9*). Therefore, new methods of isotope enrichment, with high selectivity and throughput, but also less harmful environmental effects, are desirable.

Here we report a simple and universal liquid centrifuge method to separate isotopes for almost all elements, which is at or near ambient temperature and pressure, and does not involve hazardous materials. In general, a chemical containing the element requiring isotope separation (e.g., a salt) is dissolved in a solvent (e.g., water). The solution is then centrifuged, and the heavier isotopes are enriched at the outermost portions by centrifugal force, whereas the lighter isotopes are enriched at the innermost edge (Fig. 1). The process minimizes total free energy by forming a density gradient.

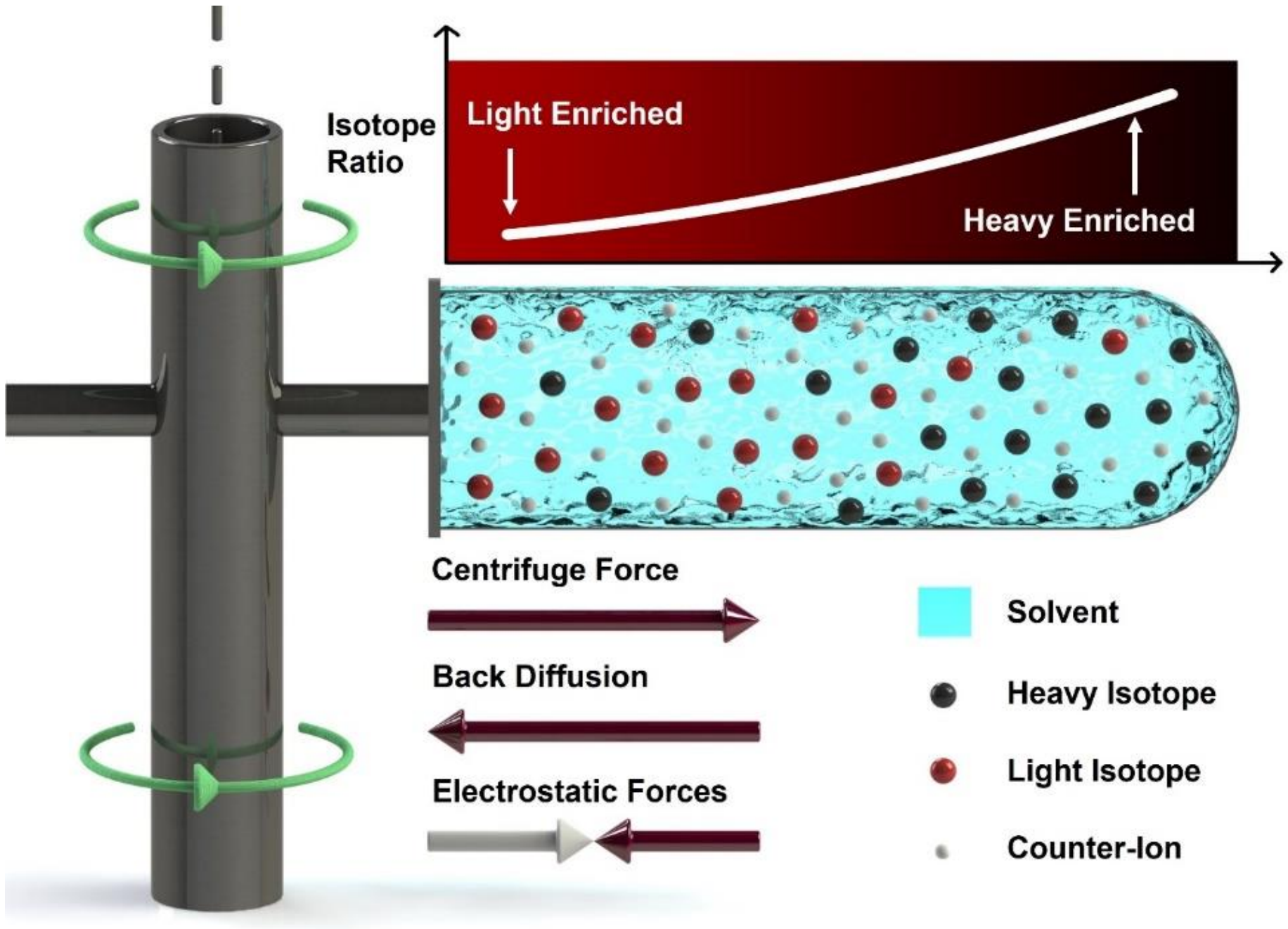

**Figure 1.** Schematic of the liquid solution centrifugation process. Heavier isotopes become more concentrated at the outer edge of the centrifuge while lighter isotopes become more concentrated at the inner edge. The electrostatic force acts to attract the target ions and counter-ions to one another, resulting in charge neutrality.

The centrifuge process is described by Equation 1 and explained further in Sections S2-3 of the Supporting Information.

$$J_i = -D_i \vartheta \frac{\partial c_i}{\partial r} + D_i \frac{\omega^2 r}{RT} c_i M_i (1 - \bar{v}_i \rho_{soln}) + D_i c_i \frac{z_i F E}{RT} \quad with \quad \vartheta = 1 + c \frac{\partial ln(\gamma)}{\partial c} \qquad (1)$$

, where $J$ is the species flux and '$i$' indexes an ionic isotope. $D$ is the species diffusivity, $\vartheta$ is the thermodynamic factor of the Onsager-Fuoss model, which relates the diffusivity, $D_F$, to the purely kinetic diffusivity $D_i$ through $D_F = \vartheta D_i$ (*10*). $c$ is the molar concentration, $r$ is the radius, $\omega$ is the angular velocity, $M$ is the molar mass, $R$ is the gas constant, and $T$ temperature. $\bar{v}$ is the partial specific volume and $\rho_{soln}$ is the density of the solution. $z$ is the valence of the ionic species, $F$ is the Faraday constant, and $E$ is the electric field. $\gamma$ is the activity coefficient.

At equilibrium, the diffusion flux due to the concentration gradient balances the mass-dependent flux arising from centrifugation as well as the electrostatic flux.

The selectivity, α, which is defined as $\alpha = ([M_1]/[M_2])_{inner} / ([M_1]/[M_2])_{outer}$ with $M_1<M_2$, can be expressed in equilibrium as

$$\alpha = exp\left(\frac{\omega^2(M_2 - M_1)(r_o^2 - r_i^2)}{2\vartheta RT}\right) \qquad (2)$$

where $r_o$ and $r_i$ are the outer and inner centrifuge radii, respectively. The selectivity equation is essentially identical to the gas centrifuge case, with the only difference being $\vartheta$ to account for non-idealities in the liquid solution (*11*). This equation has been previously applied to explain isotope fractionations observed in solid and molten metals upon centrifugation at elevated temperatures of >200°C (*12-14*). As shown in Fig. 2a, α can reach 1.05-1.1 per neutron difference at equilibrium at a practical rotation speed (e.g., 50-100,000 revolutions per minute / RPM), which is equivalent to 1.48-2.14 for $^{40}Ca/^{48}Ca$.

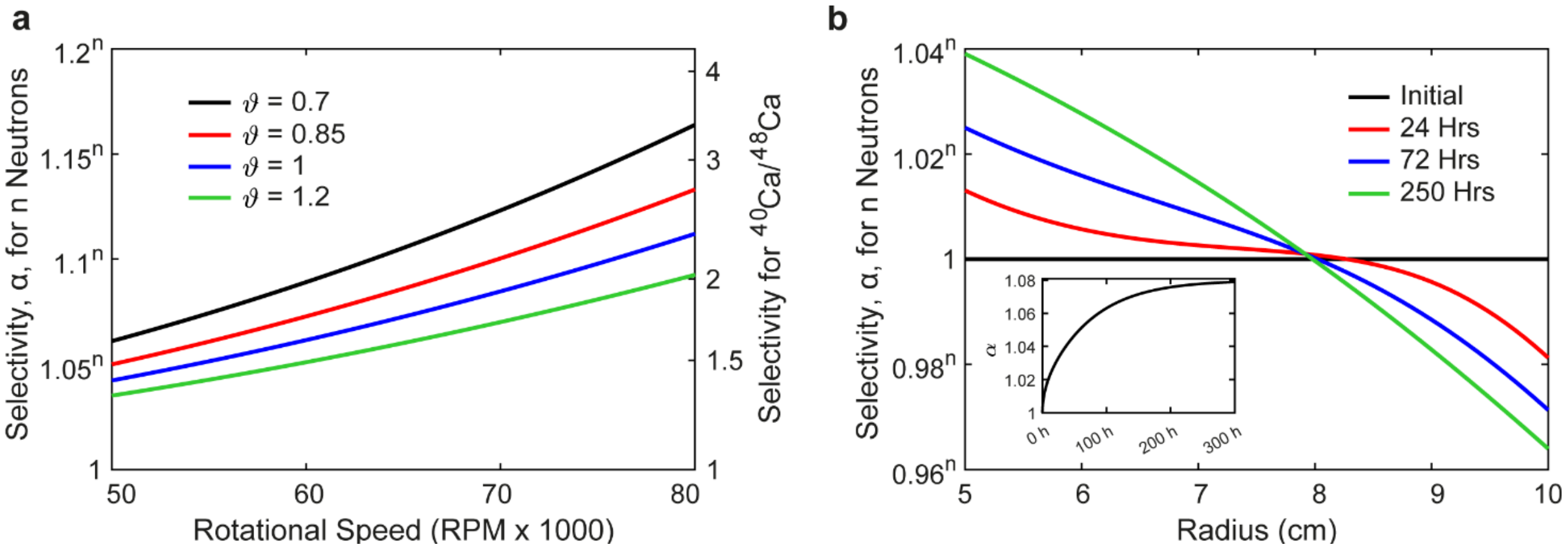

**Figure 2.** (a) Equilibrium separation factors vs RPM for a rotor with $r_i$ = 50 mm, $r_o$ = 100 mm and at 298 K. Left axis for a given element where 'n' designates the neutrons between chosen isotopes. Right axis for the case of $^{40}Ca/^{48}Ca$. (b) Time-dependent isotope ratio vs. location for 70 kRPM, 40°C, and a 1:1 monovalent salt used in the simulation. $r_i$ = 50 mm, $r_o$ = 100 mm, $D_+ = D_- = 10^{-9}$ m$^2$ s$^{-1}$, $\vartheta$ = 1. The inset figure gives the total selectivity per unit mass difference against time.

Compared to gas centrifuges, the liquid centrifuge has the following advantages. First, it is suitable for most elements while the pool for gas centrifuges is limited. For example, most elements which are difficult to form gaseous species near ambient conditions, such as all Group I and II elements and the lanthanides, are incompatible with the gas centrifuge. In contrast, every element can be made to have good water-solubility, except for the noble gases. Second, the isotope concentration in a liquid solution can be much higher than a gas. For instance, 1 mol $L^{-1}$ isotope solution is 22.4 times as concentrated as a gas at standard conditions, thus increasing throughput. Many elements can be dissolved with a concentration up to 5-10 mol $L^{-1}$ via a nitrate, nitrite, or halide (*15*). Third, $\vartheta$ is a factor that tends to be below 1 for most aqueous solutions at low concentrations and can become as low as 0.2-0.3 for certain salt/solvent combinations, particularly multivalent ions and low-dielectric-constant solvents (*16*), while $\vartheta$ is exactly 1 in an ideal gas (Section S2.1). In principle this could allow for the separation factor to be multiple times greater than a gas centrifuge with the same experimental parameters, albeit at low concentrations. Finally, non-reactive solids and liquid solutions are much safer to handle than toxic gases (e.g., $UF_6$), which have unfortunately led to fatal accidents (*17*).

Here we focus on enriching $^{48}Ca$ to demonstrate the capability of liquid centrifugation and use $^{100}Mo$ and $^{6}Li$ for further validation to represent broad classes of elements across the periodic table. $^{48}Ca$ has a natural abundance of 0.187%, while $^{40}Ca$ accounts for 96.941% of all Ca isotopes. Calcium has no suitable compound that can be gasified near ambient temperature and is currently produced by EMIS with a low rate of ~10 grams/year and price exceeding $100,000/gram. $^{100}Mo$ has a natural abundance of 9.74% and has important radiopharmaceutical applications (*18*). The current production method of $^{100}Mo$ is either low throughput (EMIS) or involves toxic chemicals such as $MoF_6$ in gas centrifugation. $^{6}Li$ has a natural abundance of 7.5% and its historical enrichment used over 2 tons of toxic Hg to obtain every 1 kg of enriched $^{6}Li$ via the COLEX process (*19*). In this report, we achieved a high selectivity of 1.434 for $^{40}Ca/^{48}Ca$ at 60 kRPM with a commercial biomedical centrifuge after 72 hours, while literature has only reported 1.005–1.012 in chemical separation (*20, 21*) and 1.26 in 14 day-long thermal diffusion (*22*). Similarly, a selectivity of 1.054 was realized in $^{6}Li/^{7}Li$, essentially identical to the COLEX process selectivity, while 1.485 was achieved in $^{92}Mo/^{100}Mo$.

To enrich $^{48}Ca$ by liquid centrifugation, $CaCl_2$, $Ca(NO_3)_2$, and $CaS_2O_3$ were dissolved in water to form 0.1 mol, 1 mol, 2 mol, or 5 mol $kg^{-1}$ solutions, which were centrifuged in a tube for 24 or 72 hours at 40°C. Samples were then taken from the top and bottom of the tubes and were analyzed with a *Nu Instruments Sapphire* collision-cell-equipped MC-ICPMS with an ultrahigh accuracy corresponding to < ±0.00065 selectivity measurement error (*23, 24*). The top and bottom of the tubes correspond to the inner and outer radii, respectively.

First, all salts tend to concentrate at the outer radii since they are denser than water, and the results are consistent with modeling predictions (Tables S13-14 and Fig. 3a). Relatively flat concentration gradients at ≥5 mol $kg^{-1}$ likely result from the theoretically understood non-linearities of transport of concentrated electrolytes. The different degrees of concentration polarization mainly originate from the magnitude of $M_{salt}(1 - \bar{v}_{salt} \rho_{soln})$, which represents the centrifugal driving force to induce such polarizations. The different diffusivities and $\vartheta$ also affect the result, but to a lesser magnitude. For example, a higher concentration of salt leads to smaller polarization due to a reduced diffusivity.

As with concentration polarizations, heavier $^{48}Ca$ also concentrates at the outer radii compared to $^{40}Ca$ (Fig. 3b). Upon time, $\alpha$ increases from 1.202 / 1.201 / 1.210 at 24 h to 1.434 / 1.410 / 1.400 at 72 h for 0.1 M $CaCl_2$, $Ca(NO_3)_2$ and $CaS_2O_3$, respectively, which is consistent with the model prediction that the counterion does not significantly affect $\alpha$. $\alpha$ generally decreases with increasing concentration as the reduced diffusivity slows down the separation before reaching equilibrium. For example, in $CaCl_2$, $\alpha$ decreases from 1.202 / 1.434 at 0.1 M to 1.162 / 1.352 at 2 M to 1.106 / 1.194 at 5 M for 24 h / 72 h, respectively (Fig. S2). However, as the enriched isotope flux is proportional to salt concentration, a higher concentration often favors a larger throughput for practical applications. The diffusivity can also be enhanced by increasing temperature, as it increases by ~2.5% / K in aqueous solution, though this will slightly decrease the equilibrium selectivity due to the $T^{-1}$ dependence of Equation 2 (*25*).

Along with the separation factor, the symmetry of isotope enrichment at the two ends of the centrifuge relative to the initial isotope ratio also plays an important role in the separation. With the same $\alpha$, if enrichment of the heavier isotope is targeted (e.g., $^{48}Ca$), a lower $([M_1]/[M_2])_{outer} / ([M_1]/[M_2])_{initial}$ ($\alpha_{outer}$) is preferred if $M_1<M_2$. If the target isotope is the lighter one (e.g., $^{10}B$), a higher $([M_1]/[M_2])_{inner} / ([M_1]/[M_2])_{initial}$ ($\alpha_{inner}$) is preferred. As shown in Fig. 3b, $\alpha_{outer}$ and $\alpha_{inner}$ are approximately symmetric within 24 hours, but with increasing time, $\alpha_{outer}$ starts to saturate while $\alpha_{inner}$ keeps increasing. $CaCl_2$ tends to have the highest $\alpha_{outer}$ and it was found that this can be mainly attributed to the value of $M_{salt}(1 - \bar{v}_{salt} \rho_{soln})$, with a smaller value favoring a higher proportion of enrichment at the outer radii.

Ideally a large proportion of the solution would have high isotopic enrichment. To determine the spatial distribution, 10 wt.% gelatin was added in the aqueous solution and the temperature was decreased to 0°C for the last 3 hours of centrifugation. The solution would mostly gelatinize, and the spatial distribution could be determined without being disturbed by convection. It is found that the heavy isotope is enriched at the bottom ~1/4 of the tube, which is consistent with model predictions for the calcium nitrate salt (Fig. 3c). However, as the salt is more concentrated at the bottom, ~1/3 of the total salt is enriched with $^{48}Ca$, and $^{40}Ca/^{48}Ca$ is roughly linear with atomic percentage in the enriched region.

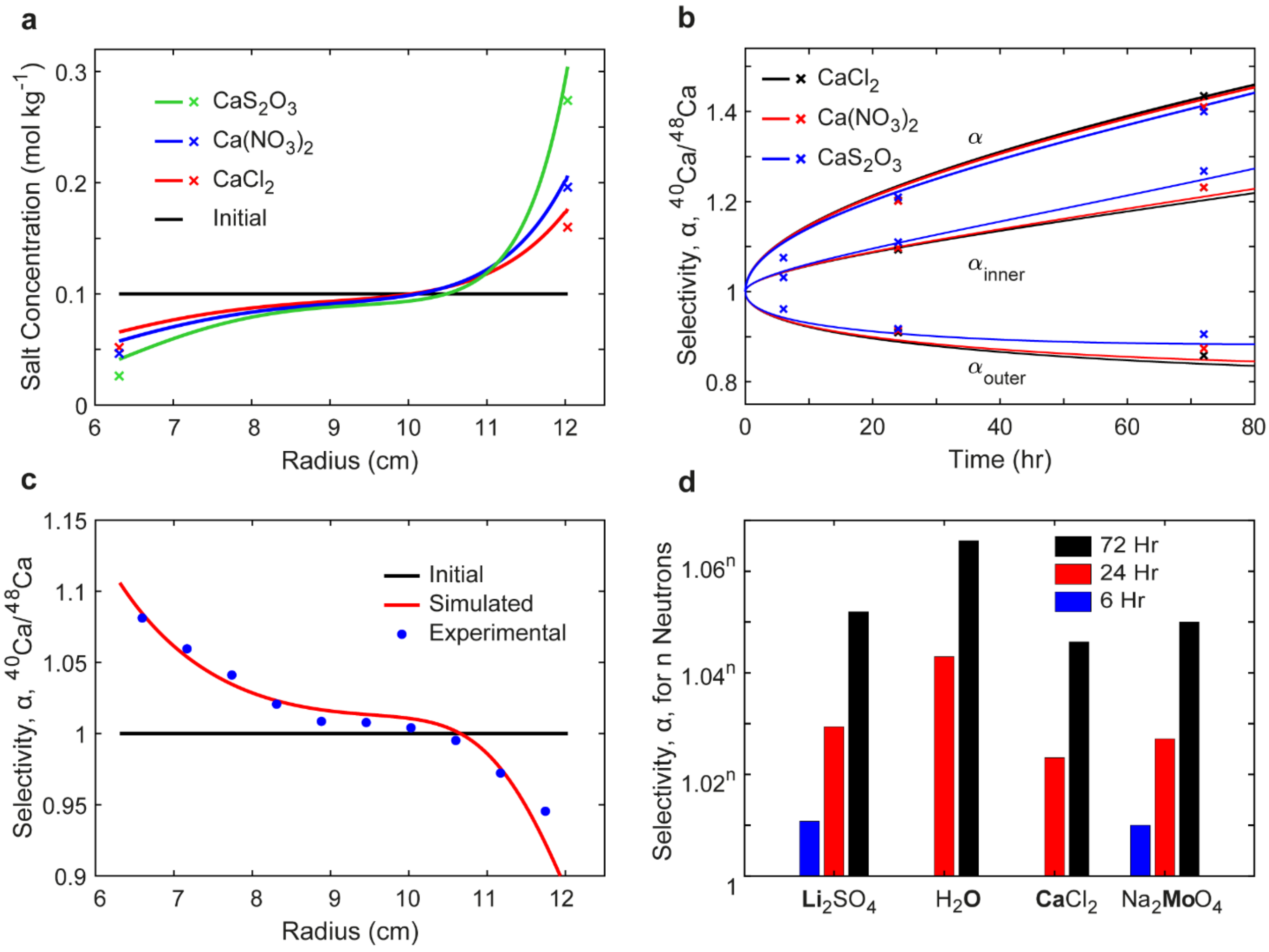

**Figure 3.** (a) Simulated concentration distributions (lines) for three calcium salts after 24 hours along with experimental values (crosses) at the inner and outer radii. (b) Simulated and experimental separation factors versus time for three Ca salts at 0.1 mol kg$^{-1}$. (c) The simulated and experimental $^{40}Ca/^{48}Ca$ selectivity spatial distribution of $Ca(NO_3)_2$ after 24 hours. (d) The measured separation factor after 6, 24, and 72 hours in selected bold elements at 0.1 mol kg$^{-1}$ in $H_2O$. Variations between elements are primarily caused by their different (ionic) diffusivities. The results are normalized to the neutron difference, n.

To further show that the liquid centrifuge method is universal across the periodic table, we applied the same method to $^{6}Li/^{7}Li$ and all seven Mo isotopes. In many cases, the same salt was used to separate the isotopes of both the anion and cation, for example $Li_2MoO_4$. As shown in Fig. 3d, $Li_2SO_4$ gives $\alpha$ of 1.052 for $^{6}Li/^{7}Li$ at 72 hours while $Na_2MoO_4$ gives an $\alpha$ of 1.485 for $^{92}Mo/^{100}Mo$, which corresponds to 1.0507 per unit mass difference. Such data confirms that $\alpha$ scales with $\Delta M$, and demonstrates the universality of the liquid centrifuge method. This is further confirmed by comparing $\alpha$ among different isotope pairs in molybdenum (Section S4). Moreover, by analyzing $^{1}H/^{2}H$ and $^{16}O/^{18}O$ in the solvent, separation factors of 1.067 and 1.134 were found, respectively, thereby indicating that isotopes within the solvent itself were effectively separated. The high self-diffusivity of water and its isotopologues means that 72 hours is sufficient to closely approach equilibrium, thereby explaining the higher value per unit mass difference of 1.065-1.067 compared to that for dissolved ions.

To demonstrate the simple scalability of liquid solution centrifugation, we performed a three stage enrichment of $^{48}Ca$. Each solution was centrifuged for 24 hours and then the top and bottom 10% of the solution was collected. These solutions were then diluted and centrifuged for an additional 24 hours, and the process repeated. The dilution step would not be required in a continuous process since the exit stream of one stage would directly feed into another. The results of these enrichments are shown in Table S12, with the $^{48}Ca$ abundance reaching 0.233 at% after the three-stage enrichment, compared to its natural abundance of 0.187 at%.

The scalability of liquid centrifugation to a continuous process is supported by analogy with the widely used gas centrifuge method and associated countercurrent devices. In this case, the feed and product streams are continuous, while the centrifuge continues to rotate at the target speed. Internal flow profiles are induced which lead to much larger separation factors along the axial direction than can be achieved in the radial direction alone. This has the effect of multiplying the single-stage separation factor by many times depending on the height/diameter ratio, such that even under modest centrifuging conditions selectivities exceeding 1.20 per unit mass difference can be achieved (e.g., >4.30 in $^{40}Ca/^{48}Ca$, >1.44 in $^{16}O/^{18}O$). This countercurrent centrifugation method greatly simplifies the cascading process by reducing the number of stages and has successfully led to the enrichment of 1000's of tons of $^{235}U$. These principles are further discussed in Section S7.

One potential challenge for the technique is that the diffusivity of ions in water is typically 1-2 $\times 10^{-9}$ $m^2$ $s^{-1}$ at 25°C, which is about three orders of magnitude less than gases for centrifugation (~$10^{-6}$ $m^2$ $s^{-1}$, Table S15). At steady state, the isotope flux is proportional to diffusivity. However, this can be largely compensated by the higher concentration of isotopes in a liquid, which is commonly >$10^2$ times that of a gas for dissolved salts, and $10^3$-$10^4$ times for solvent isotopes since the isotope flux is also proportional to its concentration (Section S8). Moreover, steady state fluxes can be increased with temperature since the diffusivity, and generally the salt solubility, increase significantly with temperature. This dependence is much stronger in a liquid than in a gas (Fig. S4). Therefore, liquid centrifugation can achieve steady state fluxes around ~1/10 that of gas centrifuges at ambient conditions, with elevated temperatures (60-100°C) allowing additional increases of 2-3 times.

Using the analogy of gas centrifuges and their available cost data for separative work, the production costs may be approximated for a liquid solution (*26, 27*). The price per mole of Separative Work (MSW) is estimated as \$70-\$120 / ΔM for a given element, where ΔM is the neutron difference between the isotopes of interest. This is in comparison to ~\$71 / ΔM per mole of separative work for $UF_6$ in operational gas centrifuges. Analysis indicates ~\$15/MSW for $^{40}Ca/^{48}Ca$. In general, the variations between elements depends on the maximum isotope fluxes and ΔM, with larger values allowing for more efficient separation. Further details are provided in Section S9.

The overall effectiveness and efficiency of this technique could be improved in real systems through several considerations, such as increasing the peripheral speed, using countercurrent centrifugation, increasing temperature, and adjusting $r_i$ and $r_o$. Finally, the isotopes of multiple desired elements can be separated at the same time by using carefully chosen salts, such as $Li_2MoO_4$ or $CaCl_2$ to simultaneously separate both Li and Mo or Ca and Cl, respectively. Moreover, isotopes in the solvent are separated and enriched simultaneously in the liquid centrifuge. Therefore, $^{18}O$-enriched water can be produced at the same time as $^2H$, $^{37}Cl$, and $^{48}Ca$ with a high concentration of ~50-55 mol $L^{-1}$, which is important for producing $^{18}F$ for positron emission tomography (*28*).

An additional factor to explore in liquid centrifugation is the solvent. Based on the Debye-Hückel theory, a solvent with lower dielectric constant (e.g., organic solvents) could reduce $\vartheta$ to 0.3-0.5 at low concentrations, which could increase the selectivity by 50-200% (*29*). However, these solvents often result in an ionic diffusivity one to two orders of magnitude lower than water, and so the transient selectivities in organic solvents in our preliminary studies were lower than in water (Table S11). However, it remains a possibility that organic solvents with low viscosities and large non-idealities could provide better results than water in some cases, including cases where the target isotope is inside these solvents (Section S10).

In general, at moderate centrifuge speeds and radii, new single stage separation factor benchmarks could be set for the majority of elements. Most importantly, this requires no adaptations to the centrifugation setup in any cases, and therefore developments of the technique and hardware for a particular elemental system, for example improvements to the centrifuge materials, represent progress for all elements. This method has potential to improve the supply of hundreds of stable isotopes which are used in various areas of energy generation, fundamental science, and radiopharmaceuticals, thereby aiding many of the important questions throughout fields of science.

**References and Notes:**

1. Meeting Isotope Needs and Capturing Opportunities for the Future: The 2015 Long Range Plan for the DOE-NP Isotope Progarm, *NSAC Isotopes Subcommitee, July 2015* (2015).
2. J. W. Beams, L. B. Snoddy, A. R. Kuhlthau, Tests of the theory of isotope separation by centrifuging. 2nd U.N. Conference on the Peaceful Uses of Atomic Energy. **4**, 428-434 (1958).
3. L. O. Love, Electromagnetic Separation of Isotopes at Oak Ridge. *Science.* **182**, 10 (1973).
4. T. Graham, On the molecular mobility of gases. *Phil. Trans. R. Soc* **153**, 385-405 (1863).
5. H. C. Urey, The thermodynamic properties of isotopic substances. *J. Chem. Soc.* **1**, 562-581 (1947).
6. V. S. Letokhov, Laser isotope separation. *Nature.* **277**, 605-610 (1979).
7. R. L. Murray, K. E. Holbert, *Nuclear Energy (Eighth Edition), Chapter 15 - Isotope Separators*. (Wiley, 2020).
8. A. N. Cheltsov, L. Y. Sosnin, V. K. Khamylov, Centrifugal enrichment of nickel isotopes and their application to the development of new technologies. *J Radioanal Nucl Chem*. **299**, 981–987 (2014).
9. J. Bigeleisen, M. G. Mayer, Calculation of Equilibrium Constants for Isotopic Exchange Reactions. *J. Chem. Phys* **15**, (1947).
10. L. Onsager, R. M. Fuoss, Irreversible processes in electrolytes. diffusion, conductance and viscous flow in arbitrary mixtures of strong electrolytes. *J. Phys. Chem.* **36**, 2689-2778 (1932).
11. H. W. Hsu, [Chapter III] in *Separations by centrifugal phenomena*. (Wiley, 1981). pp. 50-54.
12. T. Osawa, M. Ono, F. Esaka, S. Okayasu, Y. Iguchi, T. Hao, M. Magara, T. Mashimo. Mass-dependent isotopic fractionation of a solid tin under a strong gravitational field. *EPL.* **85**, 6, 64001 (2009).
13. T. Mashimo, M. Ono, X. Huang, Y. Iguchi, S. Okayasu, K. Kobayashi, E. Nakamura, Isotope separation by condensed matter centrifugation: Sedimentation of isotope atoms in Se. *J. Nucl. Sci. Technol.* **45**, 6, 105-107 (2014).

14. M. Ono, T. Mashimo, Sedimentation process for atoms in a Bi-Sb system alloy under a strong gravitational field: A new type of diffusion of substitutional solutes. *Philos. Mag A.* **82**, 3, 591-600 (2002).
15. X. Ge, X. Wang, M. Zhang, S. Seetharaman, Correlation and prediction of activity and Osmotic coefficients of aqueous electrolytes at 298.15 K by the modified TCPC model. *J. Chem. Eng. Data.* **52**, 2, 538-547 (2007).
16. N. Xin, Y. Sun, C. J. Radke, J. M. Prausnitz, Osmotic and activity coefficients for five lithium salts in three non-aqueous solvents. *J. Chem. Thermodyn.* **132**, 83-92 (2019).
17. D. Brugge, J. L. deLemos, C. Bui, The Sequoyah Corporation fuels release and the church rock spill: Unpublicized nuclear releases in American Indian Communities. *Am. J. Public Health.* **97**, 9, 1595-1600 (2007).
18. B. L. Zaret, F. J. Wackers, Nuclear Cardiology. *N Engl J Med.* **329**, (1993).
19. A. A. Palko, J. S. Drury, G. M. Begun, Lithium isotope separation factors of some two-phase equilibrium systems. *J. Chem. Phys.* **64**, 1828-1837 (1976).
20. D. Zucker, J. S. Drury, Separation of calcium isotopes in an Amalgam system. *J. Chem. Phys.* **41**, 1678-1681 (1964).
21. A. Rittirong, T. Yoshimoto, R. Hazama1, T. Kishimoto, T. Fujii, Y. Sakuma, S. Fukutani, Y. Shibahara, A. Sunaga, Isotope separation by DC18C6 crown-ether for neutrinoless double beta decay of $^{48}$Ca. *J. Phys.: Conf. Ser.* **2147**, (2022).
22. W. M. Rutherford, K. W. Laughlin, Separation of Calcium Isotopes by Liquid Phase Thermal Diffusion. *Science.* **211**, 1054-1056 (1981).
23. W. Dai, F. Moynier, M. Paquet, J. Moureau, B. Debret, J. Siebert, Y. Gerard, Y. Zhao, Calcium isotope measurements using a collision cell (CC)-MC-ICP-MS. *Chem. Geol.* **590**, (2022).
24. H. Chen, N. J. Saunders, M. Jerram, A. N. Halliday, High-precision potassium isotopic measurements by collision cell equipped MC-ICPMS. *Chem. Geol.* **578**, (2021).
25. E. A. Hollingshead, A. R. Gordon, The Differential Diffusion Constant of Calcium Chloride in Aqueous Solution. *J. Chem. Phys.* **9**, (1941).
26. Uranium Marketing Annual Report United States Energy Information Administration (EIA), (2022).
27. G. Rothwell, Market Power in Uranium Enrichment. *Sci. Glob. Secur.* **17**, 132-154 (2009).
28. M. Blau, R. Ganatra, M. A. Bender, 18F-fluoride for bone imaging. *Seminars in Nuclear Medicine.* **2**, (1972).
29. J. L. Liu, C. L. Li, A generalized Debye-Hückel theory of electrolyte solutions. *AIP Adv.* **9**, (2019).
30. F. A. Lindemann, F. W. Aston, The possibility of separating isotopes. *Phil. Mag.* **37**, 523-534 (1919).
31. G. J. Hooyman, *"Thermodynamics of Diffusion and Sedimentation" in Ultracentrifugal Analysis in Theory and Experiment*. (J. W. Williams, Ed. (Academic Press, 1963), 1962), pp. 3-12.
32. M. Brown, E. G. Murphy, Measurements of the self-diffusion coefficient of uranium hexafluoride. *Trans. Faraday Soc.* **61**, 2442-2446 (1965).

**Acknowledgements:**

We thank the Columbia Precision Biomolecular Characterization Facility (PBCF) and Dr. J. Ma for our use of their centrifuge. **Funding:** This work was supported by the United States Department of Energy, Grant Number DE-SC0022256, and the seed funding support from Columbia University's Research Initiatives in Science & Engineering (RISE) competition, started in 2004 to trigger high-risk, high-reward, and innovative collaborations in the basic sciences, engineering, and medicine. **Author Contributions:** Y.Y. initiated the studies. J.F.W. conceived the concept. H.C., J.F.W., K.L., and S.E.C. performed experimental measurements. All authors contributed to experiment planning and data analysis. J.F.W., Y.Y., and J.L. formulated the model and performed the simulations. J.F.W., K.L., H.C., A.N.H., and Y.Y. wrote the manuscript. Y.Y. and A.N.H. supervised the work. **Declaration of Interests:** A provisional patent (U.S. 63/425,181) has been filed related to this work. **Data and materials availability:** All data are available in the manuscript or the supplementary materials. Information requests should be directed to the corresponding author.

**Supplementary Materials:**

Materials and Methods

Supplementary Text

Figs. S1 to S4

Tables S1 to S17

References (*30*-*32*)

Supplementary Materials for

# Liquid solution centrifugation for safe, scalable, and efficient isotope separation

Joseph F. Wild[1*], Heng Chen[2*], Keyue Liang[1], Jiayu Liu[1], Stephen E. Cox[2], Alex N. Halliday[2†], Yuan Yang[1†]

[1]Department of Applied Physics and Applied Mathematics, Columbia University, New York, NY 10027, United States

[2]Lamont-Doherty Earth Observatory, Columbia University, Palisades, NY 10964, United States

*These authors contributed equally to this work

[†]Corresponding author. Email: yy2664@columbia.edu, alexhalliday@climate.columbia.edu

This file includes

## Section S1. Materials and Methods

**1.1. Preparing Solutions:** All chemicals used are listed in Table S1. All solutions to be centrifuged were prepared in 10 g of deionized water (Direct-Q 3 UV water purification system). E.g., 5 mol $kg^{-1}$ LiCl was prepared by adding 0.05 moles (2.12 g) of anhydrous LiCl to 10 g water. If the salt was initially hydrated, the mass of the water in the hydrated salt was subtracted from the 10 g of water. E.g., 2 mol $kg^{-1}$ $CaCl_2$ was prepared by adding 0.02 moles (2.94 g) of $CaCl_2{\cdot}2H_2O$ to 9.28 g water, since 0.72 g water was already in the hydrated salt. Solutions were prepared in 22 mL polypropylene vials which had been cleaned with deionized water and ethanol to avoid ion contamination from vials. Each centrifuge tube has a volume of around 4.0 mL. Two centrifuge tubes were used for each solution to check repeatability, and these were placed on opposite sides of the rotor after ensuring equal masses for stability.

**1.2. Centrifugation:** The SW 60 Ti rotor in the Beckman Optima XPN-100 Ultracentrifuge was used at 60,000 RPM for all experiments. The inner and outer radii are 63.1 mm and 120.3 mm, respectively. It took 4-5 minutes to reach 60,000 RPM, or 0 RPM at the end of the run. The centrifuge automatically engaged its vacuum system when the rotor reached 3,000 RPM. The rotor was generally initially at 15-20°C upon starting centrifugation and the heating rate was found to be around 0.4 - 0.5°C $min^{-1}$, so it would take around 1 hour to reach 40°C. At the end of the run, the temperature was set to 25°C for 1 hour at the same speed to bring the solution closer to ambient conditions and minimize convection-induced remixing upon collection. Open-top thinwall polypropylene tubes were used in all experiments.

**1.3. Sample Collection:** 0.5 mm sterile needles were used to collect the samples from the top and bottom of the centrifuge tubes immediately after the end of the run. This process would take around 10 minutes for all six tubes. Generally, 25-75 mg of the sample was collected in each case. The mass of the collected samples was measured by calculating the difference between the mass of the sample container before and after collection to 0.1 mg. This allowed for the concentration to be later determined. The top liquid could be accessed at the top of the centrifuge tube, while the bottom liquid was accessed by carefully removing the thinwall tubes from the bucket and then slowly piercing the bottom of the tube in a twisting motion.

**1.4. Isotopic and Concentration Measurements:** A *Nu Sapphire* MC-ICPMS (SP004; equipped with collision cell) was used for all Ca concentration and isotope measurements to minimize spectral interference from the argon support gas fueling the plasma. A separate *Nu Sapphire* MC-ICPMS (SP005 without collision cell) was used for all Li and Mo measurements. A Picarro L2130i was used for the water H and O isotope measurements. The collected solutions, as well as the references, were diluted to 50-300 ppb in 2 wt.% nitric acid. Once the concentration measurements had been made by comparing the intensities of the top and bottom solutions to the reference, the concentrations of all samples were brought within 10% of the standard solution concentration. Isotopic measurements were then made three times for each sample, and each analysis consists of 40 (50 for Li) cycles of 4s (3s for Li) integrations. Before isotopic measurements, the calcium samples were first passed through chromatography columns filled with Sr-Spec resin to remove interference element of Sr. Every isotope sample and NIST (National Institute of Standards and Technology) standard solution for that element were measured alternately (sample-standard bracketing) for mass bias and signal drift correction (*23*, *24*).

**1.5. Materials Used:**

**Table S1**

| Chemical | Source | Notes |
|---|---|---|
| calcium chloride dihydrate, ⩾99.0% | Sigma, C3306 | Lot: SLBZ8395 |
| calcium chromate, 99.9% | Alfa Aesar, 43333-14 | Metals basis, Lot: R27H007 |
| calcium nitrate tetrahydrate, 99% | Sigma, 237124 | Lot: MKCQ1963 |
| calcium thiosulfate, pure 30-50% solution in water | Acros Organics, 447870010 | Liquid solution of density 1245 kg $m^{-3}$ (~1.93 M), Lot: A0412159 |
| dimethyl sulfoxide, 99.9% min | Alfa Aesar, 36480 | Lot: X17C012 |
| lithium bis(trifluoromethanesulfonyl)imide | Gotion | Stored in Ar glovebox |
| lithium bromide, anhydrous, ⩾99% | Sigma, 746479 | Lot: MKCH6662 |
| lithium chloride, anhydrous, 98+% | Alfa Aesar, A10531 | Lot: 10189192, Stored in Ar glovebox |
| lithium hydroxide monohydrate, ⩾99.0% | Sigma, 62528 | Lot: BCBZ5579 |
| lithium molybdenum oxide, (lithium molybdate), 99+%) | Thermo Fisher, 13427 | Lot: W16G020 |
| lithium nitrate | Sigma, 227986 | Lot: MKCK5638 |
| lithium oxalate, 99+% | Thermo Fisher, 13426 | Lot: R13I014 |
| lithium sulfate, ⩾98.5% | Sigma, L6375 | Lot: BCBP6592V |
| nitric acid, ICP-OES | Thermo Scientific, T00309-0500 | Lot: 201922, For trace metal analysis |
| propylene carbonate, anhydrous, 99.7% | Sigma, 310328 | Lot: SHBJ2151 |
| sodium molybdate, anhydrous, 99.9% trace metal basis | Sigma, 737860 | Lot: MKBH2923V |
| triethyl phosphate, ≥99.8% | Sigma, 538728 | Lot: MKCJ1157, Stored in Ar glovebox |

## Section S2. Separation of Neutral Species in Equilibrium

**2.1. Ideal Gases and Liquids:** With the discovery of isotopes in the 1910's, Lindemann and Aston (*30*) first proposed centrifugation as a method of separation in 1919, and derived the following equilibrium governing equation for the case of an ideal gas or incompressible ideal liquid:

$$\alpha = exp\left(\frac{\omega^2(M_2 - M_1)\left(r_o^2 - r_i^2\right)}{2RT}\right)$$

**2.2. Nonideality:** In 1963, Hooyman (*31*) applied diffusion thermodynamics to centrifugation in multicomponent systems for the following equilibrium equation:

$$\frac{M_i\omega^2 r}{RT}[1 - \bar{v}_i(r)\rho(r)] = \frac{1}{c_i(r)}\frac{dc_i(r)}{dr} + \sum_{k=1}^{v-1}\left[\frac{\partial \ln\left(\gamma_i^{(c)}\right)}{\partial c_k(r)}\right]_{T,P,c_{j\neq k}}\frac{\partial c_k(r)}{dr}$$

In the case of a dilute solution of chemically identical isotopic species, this equation becomes:

$$\frac{M_i\omega^2 r}{RT}[1 - \bar{v}_i\rho_0] = \frac{1}{c_i(r)}\left[1 + c_{tot}(r)\frac{\partial \ln(\gamma)}{\partial c_{tot}(r)}\right]\frac{dc_{tot}(r)}{dr} = \frac{\vartheta(c_{tot})}{c_i(r)}\frac{dc_i(r)}{dr}$$

If the concentration dependence of the thermodynamic factor can be neglected, then the resulting differential equation is separable and can be solved analytically:

$$\frac{M_i\omega^2 r}{RT}[1 - \bar{v}_i\rho_0]\,dr = \frac{\vartheta}{c_i(r)}dc_i(r) \quad \rightarrow \quad c_i(r) = B_i\, exp\left(\frac{\omega^2 M_i[1 - \bar{v}_i\rho_0]r^2}{2\vartheta RT}\right)$$

Then, if it is assumed that the partial molar volume of a chemical is the same for isotopes, $M_1\bar{v}_1 = M_2\bar{v}_2$. Finally, using the definition of the selectivity:

$$\alpha = \frac{c_2(r_o)/c_1(r_o)}{c_2(r_i)/c_1(r_i)} = \frac{\frac{B_2}{B_1}exp\left(\frac{\omega^2(M_2 - M_1){r_o}^2}{2\vartheta RT}\right)}{\frac{B_2}{B_1}exp\left(\frac{\omega^2(M_2 - M_1){r_i}^2}{2\vartheta RT}\right)} = exp\left(\frac{\omega^2(M_2 - M_1)\left(r_o^2 - r_i^2\right)}{2\vartheta RT}\right)$$

This equation neglects the pressure dependence of the thermodynamic factor as well as any solvation shell affects. As will be shown in Section 3, this final equation does not change if the anion and cation are treated separately, and their motion is coupled via an electric field term.

## Section S3. Ionic Species Kinetics and Thermodynamics

**3.1. Kinetic Model:** Upon the dissolution of a salt into a solvent, the ions will tend to disassociate. Each ion will then respond to an external field depending on its own physical properties, i.e., ionic mobility, mass, volume, charge, etc. Macroscopically, the anion and cation move together due to the condition of charge neutrality, and so they are coupled via their electrostatic interaction. To incorporate this into a single model, the following equations were proposed for the 1D case. In the most simplified case of a single isotopic anion and two isotopic cations, they are:

$$J_- = -D_- \vartheta \frac{\partial c_-}{\partial r} + D_- \frac{\omega^2 r}{RT} c_- M_- (1 - \bar{v}_- \rho_{soln}) + D_- c_- \frac{z_- F E}{RT} \quad \text{(S1)}$$

$$J_{+,1} = -D_+ \vartheta \frac{\partial c_{+,1}}{\partial r} + D_+ \frac{\omega^2 r}{RT} c_{+,1} M_{+,1} \left(1 - \bar{v}_{+,1} \rho_{soln}\right) + D_+ c_{+,1} \frac{z_{+,1} F E}{RT} \quad \text{(S2)}$$

$$J_{+,2} = -D_+ \vartheta \frac{\partial c_{+,2}}{\partial r} + D_+ \frac{\omega^2 r}{RT} c_{+,2} M_{+,2} \left(1 - \bar{v}_{+,2} \rho_{soln}\right) + D_+ c_{+,2} \frac{z_{+,2} F E}{RT} \quad \text{(S3)}$$

The first term on the right is the Fickian flux, the second is the centrifugal flux, and the third term is the electrostatic flux. The electrostatic term is the only one that couples the ions together.

In addition to these, there is also the conservation of mass and the boundary/initial conditions:

$$\frac{\partial c_i}{\partial t} = -\frac{\partial J_i}{\partial r} \quad with \quad J_i(r_{inner}, t) = J_i(r_{outer}, t) = 0 \quad and \quad c_i(t = 0, r) = c_{o,i}$$

In general, these equations are:

$$\vec{J_i} = -D_i \vartheta \nabla c_i + D_i \frac{\omega^2 \vec{\boldsymbol{r}}}{RT} c_i M_i (1 - \bar{v}_i \rho_{soln}) + D_i c_i \frac{z_i F \vec{\boldsymbol{E}}}{RT} \quad with \quad \frac{\partial c_i}{\partial t} = -\nabla \cdot \left(\vec{J_i}\right)$$

**3.2. Solving Equations:** The equations were solved numerically in MATLAB. The electrostatic term could be dealt with by enforcing charge neutrality in all locations and for all times, i.e. $z_- c_-(r,t) = -z_+ (c_{+,1}(r,t) + c_{+,2}(r,t))$. This electroneutrality results from the Poisson equation $\Delta\varphi = (z_- c_- + z_+ c_+) F/\varepsilon$. Since $F/\varepsilon > 10^{14}$ V m mol$^{-1}$ in water, even if $\Delta\varphi$ were of the order of 1 kV mm$^{-2}$ ($10^9$ V m$^{-2}$), then $z_- c_- + z_+ c_+ < 10^{-5}$ mol m$^{-3}$ = 10 nmol L$^{-1}$. Therefore, the solution can be treated as electrically neutral with the overall anion and cation charge balancing everywhere. This requirement can be combined with mass conservation to obtain

$$|z|_- J_-(r,t) = |z|_+ \left[J_{+,1}(r,t) + J_{+,2}(r,t)\right] \quad \text{(S4)}$$

at all r (radii) and for all t (time). By combining equations S1-S4, the time dependent concentration distribution of each species can be determined.

**3.3. Equilibrium Derivation**: The equilibrium selectivity can be derived from the above equations with the electrostatic term as follows, which reveals the same equation as at the end of Section S2. This value was also converged upon as $t \rightarrow \infty$ in the MATLAB simulation, as shown in Fig. 2b:

At equilibrium, $J_{+,1} = J_{+,2} = 0$. Then (S3) × $c_{+,1}$ - (S2) × $c_{+,2}$ leads to

$$\vartheta c_{+,1}\frac{\partial c_{+,2}}{\partial r} - \vartheta c_{+,2}\frac{\partial c_{+,1}}{\partial r} = \frac{\omega^2 r}{RT} c_{+,1}c_{+,2}\left[M_{+,2}\left(1-\bar{v}_{+,2}\rho_{soln}\right) - M_{+,1}\left(1-\bar{v}_{+,1}\rho_{soln}\right)\right]$$

Then using $M_{+,1}\bar{v}_{+,1} = M_{+,2}\bar{v}_{+,2}$ and the reverse quotient rule:

$$\frac{\partial\left(\frac{c_{+,2}}{c_{+,1}}\right)}{\partial r} = \frac{\omega^2 r}{\vartheta RT}\frac{c_{+,2}}{c_{+,1}}(M_{+,2} - M_{+,1})$$

This is then separable in $r$ and $\frac{c_{+,2}}{c_{+,1}}$ :

$$\frac{c_{+,1}}{c_{+,2}}\, d\left(\frac{c_{+,2}}{c_{+,1}}\right) = \frac{\omega^2 r}{\vartheta RT}\left(M_{+,2} - M_{+,1}\right)dr \quad \rightarrow \quad \frac{c_{+,2}}{c_{+,1}}(r) = B_o exp\left(\frac{\omega^2 (M_2 - M_1) r^2}{2\vartheta RT}\right)$$

Finally, using the definition of the selectivity:

$$\alpha = \frac{\frac{c_{+,2}}{c_{+,1}}(r_o)}{\frac{c_{+,2}}{c_{+,1}}(r_i)} = \frac{B_o exp\left(\frac{\omega^2 (M_2 - M_1) {r_o}^2}{2\vartheta RT}\right)}{B_o exp\left(\frac{\omega^2 (M_2 - M_1) {r_i}^2}{2\vartheta RT}\right)} = exp\left(\frac{\omega^2 (M_2 - M_1)\left(r_o^2 - r_i^2\right)}{2\vartheta RT}\right)$$

Therefore, the same equilibrium selectivity is obtained whether or not the electrostatic interaction is considered.

## Section S4. Mass Dependence

**4.1. Choice of Element:** To verify that the natural logarithm of the selectivity among different isotopes is strictly proportional to their mass difference, even at the transient state, the selectivities were measured among different isotope pairs of an element after 72 hours of liquid centrifugation. Molybdenum was chosen since this element has the highest number of highly abundant stable isotopes - all seven of its stable isotopes have high natural abundances between 9.19% and 24.29% and can therefore be measured to high precision. Moreover, molybdenum isotopes only interfere directly with some zirconium and ruthenium isotopes which do not tend to naturally contaminate molybdenum sources, and therefore background interferences can be effectively removed. The isotopes of molybdenum, along with their masses and natural abundances (at%), are given in Table S2.

**Table S2**

| Isotope | Mass (Da) | Natural Abundance |
|---|---|---|
| $^{92}Mo$ | 91.90681 | 14.65% |
| $^{94}Mo$ | 93.90509 | 9.19% |
| $^{95}Mo$ | 94.90584 | 15.87% |
| $^{96}Mo$ | 95.90468 | 16.67% |
| $^{97}Mo$ | 96.90602 | 9.58% |
| $^{98}Mo$ | 97.90540 | 24.29% |
| $^{100}Mo$ | 99.90748 | 9.74% |

These seven isotopes give 21 pairs of isotopes with which the mass dependence of the isotope separation can be tested. These pairs are given in Table S3, along with their neutron and mass differences.

**Table S3**

| Neutron Difference | Isotope Pair | Mass Difference (Da) |
|---|---|---|
| 1 | (94,95), (95,96), (96,97), (97,98) | 1.00075, 0.99884, 1.00134, 0.99938 |
| 2 | (92,94), (94,96), (95,97), (96,98), (98,100) | 1.99828, 1.99959, 2.00018, 2.00072, 2.00208 |
| 3 | (92,95), (94,97), (95,98), (97,100) | 2.99903, 3.00093, 2.99956, 3.00146 |
| 4 | (92,96), (94,98), (96,100) | 3.99787, 4.00031, 4.00280 |
| 5 | (92,97), (95,100) | 4.99921, 5.00164 |
| 6 | (92,98), (94,100) | 5.99859, 6.00239 |
| 8 | (92,100) | 8.00067 |

**4.2. Mass Dependence Results:** The centrifuge experiment chosen to test the mass dependence was the first tube of the 0.1 m $Na_2MoO_4$ 72 hour run at 40ºC since this produced a large separation and the $Na_2MoO_4$ was of very high purity, as given by its trace metal analysis in its certificate of analysis. Table S4 gives the measured selectivities for each pair of isotopes. The selectivity is defined as follows, where square brackets indicate concentrations:

$$\alpha = \frac{([M_1]/[M_2])_{Inner\ Radius}}{([M_1]/[M_2])_{Outer\ Radius}}$$

In agreement with the theory of Sections S2 and S3, the isotope separation factor was found to be an exponential function of the mass difference between the isotopes to very high accuracy. The coefficient of determination was 0.99999953 for $\ln(\alpha)$ being a linear function of $\Delta M$. As can be seen in Figures S1b and S1d, the MC-ICPMS precision was mostly able to discern the slight selectivity variations between isotopes of one neutron difference, owing to their slightly different mass differences in Table S3 due to nuclear binding energies. Assuming the mass dependence of the separation to be exactly in-line with theory, an upper-bound error on the selectivity measurement in this case can be given as ±0.00015.

Moreover, the results show that the isotope separation depends only on the isotope mass difference $\Delta M$, and not the relative mass difference $\Delta M/f(M_1M_2)$, unlike chemical exchange, gas diffusion, thermal diffusion, or distillation. Hence, liquid solution centrifugation is equally effective for both light and heavy elements on a per-neutron basis, whether this be $^{6}Li$ and $^{7}Li$ or $^{207}Pb$ and $^{208}Pb$.

**Table S4**

| Isotope | 92 | 94 | 95 | 96 | 97 | 98 | 100 |
|---|---|---|---|---|---|---|---|
| 92 | 1 | 1.103958 | 1.160020 | 1.218756 | 1.280595 | 1.345458 | 1.485320 |
| 94 | | 1 | 1.050783 | 1.103988 | 1.160004 | 1.218759 | 1.345450 |
| 95 | | | 1 | 1.050633 | 1.103942 | 1.159858 | 1.280426 |
| 96 | | | | 1 | 1.050740 | 1.103960 | 1.218718 |
| 97 | | | | | 1 | 1.050651 | 1.159867 |
| 98 | | | | | | 1 | 1.103951 |
| 100 | | | | | | | 1 |

Figure S1 shows plots of the isotope selectivity (Table S4) and log selectivity versus the isotope mass difference (Table S3), as well as regression lines for the data in red:

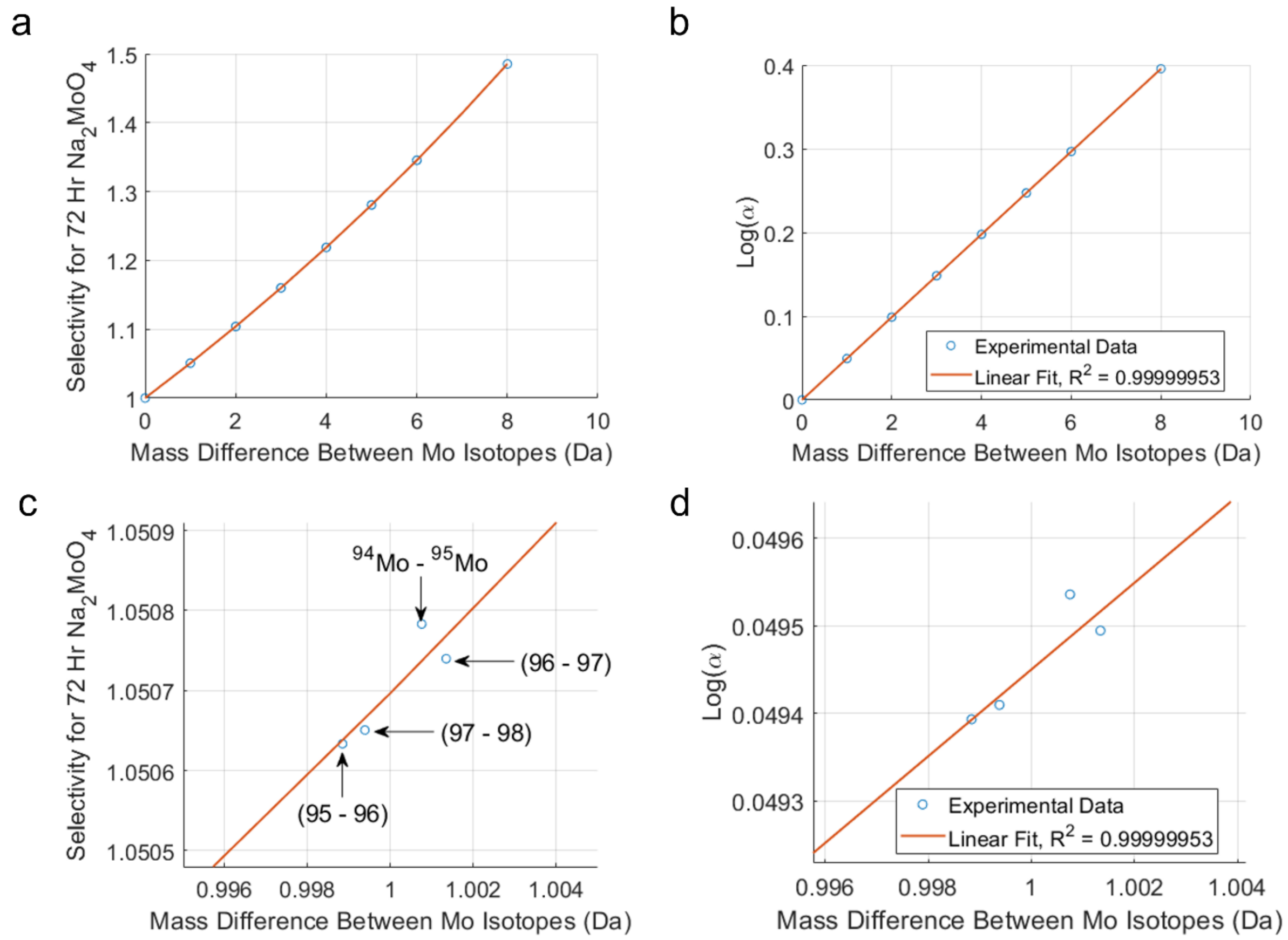

**Figure S1** - Selectivity vs Mass Difference for Mo in $Na_2MoO_4$. (c) and (d) are zoomed in versions (~1000 x) of (a) and (b), respectively, around 1 Da. (a-d) Experimental data (blue circles) and lines of best fit (red) are plotted together. The exceptional agreement indicates an MC-ICPMS precision of better than one part in $10^4$.

## Section S5. Selectivity Data

**5.1. Measurements and Errors:** As described in Section 1, two Nu Sapphire MC-ICPMS instruments were used for all isotopic measurements apart from H and O. The typical per mil errors for these measurements, given as 2x the measurement standard deviation (SD), were:

**Table S5**

| Isotope Ratio | $^{7}Li/^{6}Li$ | $^{44}Ca/^{40}Ca$ | $^{98}Mo/^{95}Mo$ | $^{18}O/^{16}O$ | D/H |
|---|---|---|---|---|---|
| 2 SD | 0.65 | 0.15 | 0.06 | 0.15 | 0.7 |

A 0.5 per mil error corresponds to a ±0.0005 selectivity error. These errors are many times smaller than the known error associated with the convection and diffusion which takes place between the centrifuge slowing down and the sample collection. Upon sample collection, a finite amount of solution is collected from the top and bottom of the centrifuge tube, and therefore some solution is taken which is not exactly at the top surface or the bottom. These last two unavoidable factors will reduce $\ln(\alpha)$ by an estimated 3-10% compared to the value just before the centrifuge begins to slow down, based on the simulation predictions and the variations between repeat values.

**5.2. Aqueous Experiments:** The results from all aqueous experiments are given in Tables S6-S10 for Ca, Mo, H, O, and Li respectively. Values for the inner selectivity, outer selectivity, and total selectivity are given, with the inner and outer selectivities defined below. 1 m salt solution = 1 mol $kg^{-1}$ water.

$$\alpha_{inner} = \frac{([M_1]/[M_2])_{Inner\ Radius}}{([M_1]/[M_2])_{Inital\ Solution}}$$

$$\alpha_{outer} = \frac{([M_1]/[M_2])_{Outer\ Radius}}{([M_1]/[M_2])_{Initial\ Solution}}$$

$$Inner\ Percentage = \ log_{\alpha}(\alpha_{inner})$$

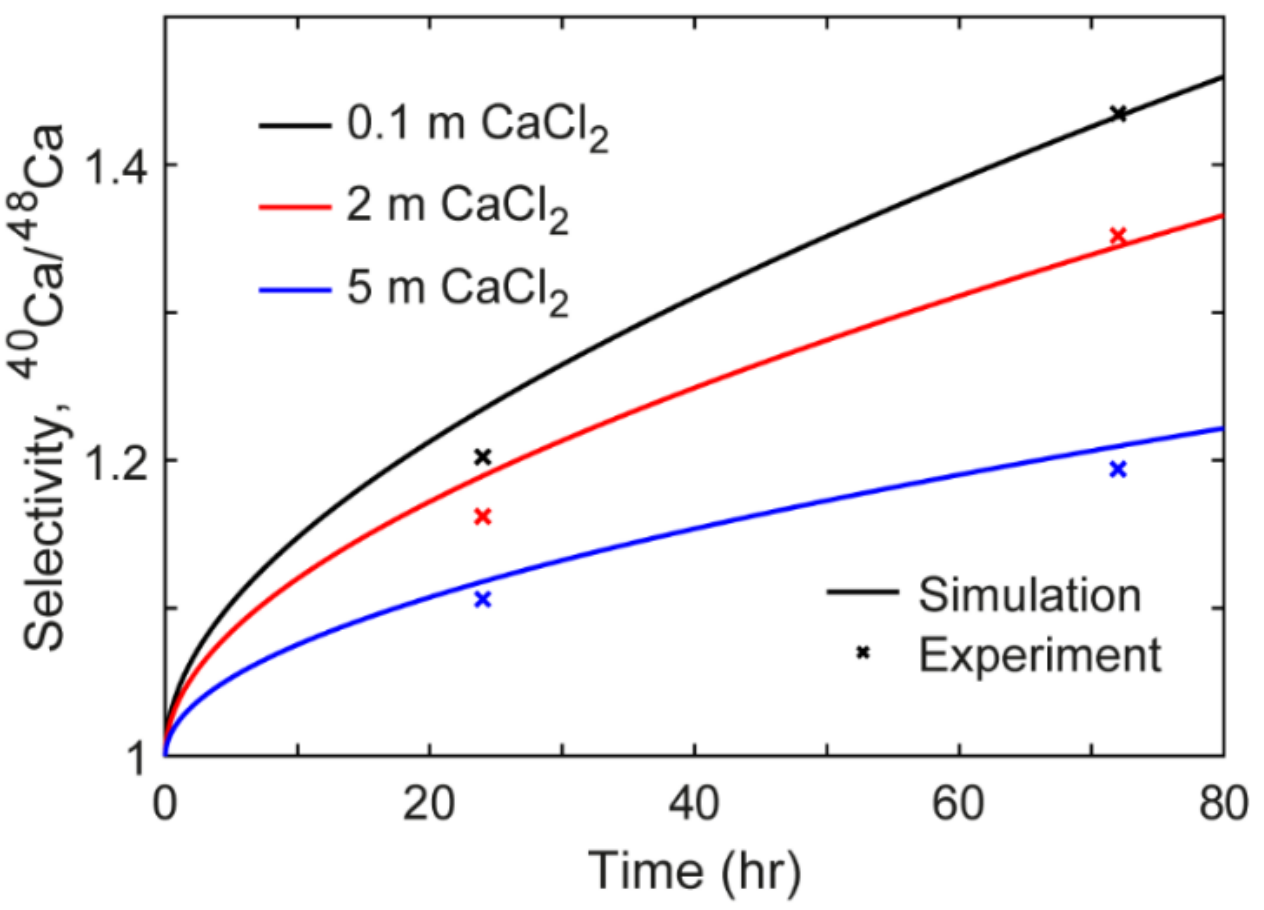

**Figure S2** - The 24- and 72-hour selectivities of 3 concentrations of $CaCl_2$. Higher concentrations give lower transient results due the lower ionic diffusivities. The larger deviations from the simulated curves after 24 hours are suspected to be caused by remixing and finite sample collection.

**Table S6 ($^{40}Ca/^{48}Ca$)**

| Sample | Time (Hours) | Inner Selectivity | Outer Selectivity | Inner Percentage | Total Selectivity |
|---|---|---|---|---|---|
| 0.1 m $CaCl_2$, 1 | **72** | 1.2312 | 0.8583 | 57.6% | 1.4345 |
| 0.1 m $CaCl_2$, 2 | **72** | 1.2167 | 0.9148* | 68.8% | 1.3300 |
| 0.1 m $Ca(NO_3)_2$, 1 | **72** | 1.2165 | 0.8726 | 59.0% | 1.3941 |
| 0.1 m $Ca(NO_3)_2$, 2 | **72** | 1.2316 | 0.8734 | 60.6% | 1.4101 |
| 0.1 m $CaS_2O_3$, 1 | **72** | 1.2267 | 0.9025 | 66.6% | 1.3592 |
| 0.1 m $CaS_2O_3$, 2 | **72** | 1.2678 | 0.9056 | 70.5% | 1.4000 |
| 2 m $CaCl_2$, 1 | **72** | 1.1616 | 0.8660 | 51.0% | 1.3413 |
| 2 m $CaCl_2$, 2 | **72** | 1.1601 | 0.8580 | 49.2% | 1.3521 |
| 2 m $Ca(NO_3)_2$, 1 | **72** | 1.1786 | 0.8776 | 55.7% | 1.3430 |
| 2 m $Ca(NO_3)_2$, 2 | **72** | 1.1731 | 0.8630 | 52.1% | 1.3583 |
| 1 m $CaS_2O_3$, 1 | **72** | 1.2140 | 0.8895 | 62.4% | 1.3648 |
| 1 m $CaS_2O_3$, 2 | **72** | 1.2323 | 0.8790 | 61.8% | 1.4019 |
| 5 m $CaCl_2$, 1 | **72** | 1.0878 | 0.9109 | 47.4% | 1.1942 |
| 5 m $CaCl_2$, 2 | **72** | 1.0850 | 0.9124 | 47.1% | 1.1892 |
| 5 m $Ca(NO_3)_2$, 1 | **72** | 1.0912 | 0.9211 | 51.5% | 1.1847 |
| 5 m $Ca(NO_3)_2$, 2 | **72** | 1.0880 | 0.9196 | 50.2% | 1.1831 |
| 0.1 m $CaCl_2$, 1 | **24** | 1.0947 | 0.9189 | 51.7% | 1.1913 |
| 0.1 m $CaCl_2$, 2 | **24** | 1.0931 | 0.9092 | 48.3% | 1.2023 |
| 0.1 m $Ca(NO_3)_2$, 1 | **24** | 1.0923 | 0.9129 | 49.2% | 1.1965 |
| 0.1 m $Ca(NO_3)_2$, 2 | **24** | 1.0974 | 0.9137 | 50.7% | 1.2011 |
| 0.1 m $CaS_2O_3$, 1 | **24** | 1.1097 | 0.9175 | 54.7% | 1.2095 |
| 0.1 m $CaS_2O_3$, 2 | **24** | 1.1067 | 0.9283 | 57.7% | 1.1922 |
| 2 m $CaCl_2$, 1 | **24** | 1.0678 | 0.9189 | 43.7% | 1.1620 |
| 2 m $CaCl_2$, 2 | **24** | - | 0.9112 | - | - |
| 5 m $CaCl_2$, 1 | **24** | 1.0412 | 0.9417 | 40.2% | 1.1057 |
| 5 m $CaCl_2$, 2 | **24** | 1.0378 | 0.9569 | 45.7% | 1.0845 |
| 0.1 m $CaS_2O_3$, 1 | **6** | 1.0317 | 0.9613 | 42.9% | 1.0754 |
| 0.1 m $CaS_2O_3$, 2 | **6** | 1.0205 | 0.9594 | 32.9% | 1.0637 |

**Table S7 ($^{92}Mo/^{100}Mo$)**

| Sample | Time (Hours) | Inner Selectivity | Outer Selectivity | Inner Percentage | Total Selectivity |
|---|---|---|---|---|---|
| 0.1 m $Na_2MoO_4$, 1 | **6** | 1.0341 | 0.9505 | 39.8% | 1.0880 |
| 0.1 m $Na_2MoO_4$, 2 | **6** | 1.0331 | 0.9526 | 39.6% | 1.0856 |
| 0.1 m $Na_2MoO_4$, 1 | **24** | 1.1227 | 0.9027 | 53.1% | 1.2437 |
| 0.1 m $Na_2MoO_4$, 2 | **24** | 1.1222 | 0.9011 | 52.5% | 1.2454 |
| 0.1 m $Na_2MoO_4$, 1 | **72** | 1.2976 | 0.8738 | 65.9% | 1.4850 |
| 0.1 m $Na_2MoO_4$, 2 | **72** | 1.2870 | 0.8756 | 65.5% | 1.4698 |

**Table S8 ($^{1}H/^{2}H$)**

| Sample | Time (Hours) | Inner Selectivity | Outer Selectivity | Inner Percentage | Total Selectivity |
|---|---|---|---|---|---|
| Water (0.5 m LiCl), 1 | **24** | 1.0198 | 0.9789 | 48.0% | 1.0417 |
| Water (0.5 m LiCl), 2 | **24** | 1.0190 | 0.9780 | 45.7% | 1.0420 |
| Water (0.5 m LiCl), 3 | **24** | 1.0191 | 0.9772 | 45.0% | 1.0429 |
| Water (0.5 m LiCl), 1 | **72** | 1.0301 | 0.9654 | 45.7% | 1.0670 |
| Water (0.5 m LiCl), 2 | **72** | 1.0297 | 0.9662 | 46.0% | 1.0657 |

**Table S9 ($^{16}O/^{18}O$)**

| Sample | Time (Hours) | Inner Selectivity | Outer Selectivity | Inner Percentage | Total Selectivity |
|---|---|---|---|---|---|
| Water (0.5 m LiCl), 1 | **24** | 1.0388 | 0.9552 | 45.4% | 1.0875 |
| Water (0.5 m LiCl), 2 | **24** | 1.0382 | 0.9541 | 44.4% | 1.0882 |
| Water (0.5 m LiCl), 3 | **24** | 1.0380 | 0.9539 | 44.2% | 1.0881 |
| Water (0.5 m LiCl), 1 | **72** | 1.0586 | 0.9331 | 45.2% | 1.1342 |
| Water (0.5 m LiCl), 2 | **72** | 1.0588 | 0.9341 | 46.6% | 1.1335 |

**Table S10 ($^6Li/^7Li$ Isotopes)**

| Sample | Time (Hours) | Inner Selectivity ($\alpha_{inner}$) | Outer Selectivity ($\alpha_{outer}$) | Inner Percentage | Total Selectivity ($\alpha$) |
|---|---|---|---|---|---|
| 0.1 m $Li_2SO_4$, 1 | **72** | 1.0274 | 0.9765 | 53.3% | 1.0520 |
| 0.1 m $Li_2SO_4$, 2 | **72** | 1.0272 | 0.9768 | 53.4% | 1.0515 |
| 1 m LiCl, 1* | **72** | 1.0241 | 0.9804 | 54.5% | 1.0447 |
| 1 m LiCl, 2 | **72** | 1.0252 | 0.9751 | 49.7% | 1.0514 |
| 1 m LiBr, 1 | **72** | 1.0290 | 0.9771 | 55.3% | 1.0531 |
| 1 m LiBr, 2 | **72** | 1.0290 | 0.9772 | 55.3% | 1.0531 |
| 1 m LiI, 1 | **72** | 1.0321 | 0.9789 | 59.6% | 1.0544 |
| 1 m LiI, 2 | **72** | 1.0316 | 0.9800 | 60.7% | 1.0526 |
| 1 m $Li_2MoO_4$, 1 | **72** | 1.0188 | 0.9800 | 48.0% | 1.0396 |
| 1 m $Li_2MoO_4$, 2 | **72** | 1.0208 | 0.9806 | 51.2% | 1.0410 |
| 2 m LiOH, 1 | **72** | 1.0208 | 0.9761 | 46.0% | 1.0458 |
| 2 m LiOH, 2 | **72** | 1.0206 | 0.9752 | 44.8% | 1.0466 |
| 5 m LiCl, 1 | **72** | 1.0198 | 0.9752 | 43.8% | 1.0458 |
| 5 m LiCl, 2 | **72** | 1.0204 | 0.9744 | 43.9% | 1.0471 |
| 0.1 m $Li_2SO_4$, 1 | **24** | 1.0129 | 0.9840 | 44.2% | 1.0294 |
| 0.1 m $Li_2SO_4$, 2 | **24** | 1.0130 | 0.9842 | 44.7% | 1.0293 |
| 0.1 m $Li_2MoO_4$, 1 | **24** | 1.0134 | 0.9863 | 49.1% | 1.0275 |
| 0.1 m $Li_2MoO_4$, 2 | **24** | 1.0135 | 0.9854 | 47.7% | 1.0285 |
| 0.1 m $Li_2C_2O_4$ ,1 | **24** | 1.0125 | 0.9843 | 43.9% | 1.0287 |
| 0.1 m $Li_2C_2O_4$, 2 | **24** | 1.0122 | 0.9853 | 45.0% | 1.0273 |
| 0.1 m LiOH, 1 | **24** | 1.0085 | 0.9879 | 40.7% | 1.0210 |
| 0.1 m LiOH, 2 | **24** | 1.0086 | 0.9888 | 43.5% | 1.0199 |
| 0.5 m $Li_2SO_4$, 1 | **24** | 1.0112 | 0.9842 | 40.9% | 1.0276 |
| 0.5 m $Li_2SO_4$, 2 | **24** | 1.0112 | 0.9845 | 41.7% | 1.0271 |
| 0.5 m $LiNO_3$, 1 | **24** | 1.0123 | 0.9832 | 42.0% | 1.0295 |
| 0.5 m $LiNO_3$, 2 | **24** | 1.0122 | 0.9823 | 40.5% | 1.0304 |
| 2 m LiOH, 1 | **24** | 1.0107 | 0.9878 | 46.6% | 1.0231 |
| 2 m LiOH, 2 | **24** | 1.0106 | 0.9873 | 45.2% | 1.0236 |
| 20% Mass LiBr, 1 | **24** | 1.0128 | 0.9863 | 48.1% | 1.0268 |
| 20% Mass LiBr, 2 | **24** | 1.0127 | 0.9854 | 46.4% | 1.0276 |
| 5 m LiCl, 1 | **24** | 1.0104 | 0.9870 | 44.2% | 1.0237 |
| 5 m LiCl, 2 | **24** | 1.0106 | 0.9869 | 44.5% | 1.0240 |
| 10 m LiCl, 1 | **24** | 1.0062 | 0.9905 | 39.2% | 1.0159 |
| 10 m LiCl, 2 | **24** | 1.0060 | 0.9905 | 38.4% | 1.0157 |
| 0.1 m $Li_2SO_4$, 1 | **6** | - | - | - | 1.0107 |
| 0.1 m $Li_2SO_4$, 2 | **6** | - | - | - | 1.0108 |

*It is suspected that accidental knocking of the centrifuged tube during collection induced convection and reduced the results.

**5.3. Non-aqueous Experiments:** Results from all non-aqueous experiments are given in Table S11.

**Table S11**

| Sample | Solvent | Time (Hours) | Inner Selectivity | Outer Selectivity | Total Selectivity |
|---|---|---|---|---|---|
| 0.1 m LiBr, 1 | **Propylene Carbonate** | **24** | - | - | 1.0054 |
| 0.1 m LiBr, 2 | **Propylene Carbonate** | **24** | - | - | 1.0054 |
| 0.1 m LiCl, 1 | **Dimethyl Sulfoxide** | **24** | - | - | 1.0056 |
| 0.1 m LiCl, 2 | **Dimethyl Sulfoxide** | **24** | - | - | 1.0050 |
| 0.1 m $LiPF_6$, 1 | **Propylene Carbonate** | **24** | - | - | 1.0048 |
| 0.1 m $LiPF_6$, 2 | **Propylene Carbonate** | **24** | - | - | 1.0050 |
| 0.1 m $LiBF_4$, 1 | **Propylene Carbonate** | **24** | - | - | 1.0045 |
| 0.1 m $LiBF_4$, 2 | **Propylene Carbonate** | **24** | - | - | 1.0046 |
| 0.2 m $Ca(NO_3)_2{\cdot}4H_2O$, 1 | **Triethyl Phosphate** | **24** | 1.0357 | 0.9715 | 1.0661 |
| 0.2 m $Ca(NO_3)_2{\cdot}4H_2O$, 2 | **Triethyl Phosphate** | **24** | 1.0451 | - | - |
| 0.2 m $Ca(NO_3)_2$, 1 | **Triethyl Phosphate** | **24** | 1.0484 | 0.9722 | 1.0784 |
| 0.2 m $Ca(NO_3)_2$, 2 | **Triethyl Phosphate** | **24** | 1.0556 | 0.9832 | 1.0736 |

**5.4. Cascade Experiments:** The results from the 3-stage 24-hour $CaCl_2$ cascade are given in Table S12. All selectivity values are given with respect to the initial (natural) $^{40}Ca/^{48}Ca$ ratio.

Two independent cascades were run with the first number of each pair of values being from the first cascade and the second being from the second cascade. The table entry above each cell shows the selectivity of the solution at the beginning of the centrifuge experiments.

**Table S12**

<table>
<tr><td>Stage</td><td colspan="6">Initial: 1.0000</td></tr>
<tr><td>1</td><td colspan="3">Top/Inner: 1.0711, 1.0744</td><td colspan="3">Bottom/Outer: 0.9188, 0.9224</td></tr>
<tr><td>2</td><td colspan="2">Top: 1.1582, 1.1578</td><td>Bottom: 0.9778, 0.9852</td><td>Top: 0.9978, 0.9982</td><td colspan="2">Bottom: 0.8396, -</td></tr>
<tr><td>3</td><td>Top: 1.2110, 1.2032</td><td>Bottom: 1.1182, 1.1326</td><td>-</td><td>-</td><td>Top: 0.8841, 0.8816</td><td>Bottom: 0.8024, 0.8023</td></tr>
</table>

## Section S6. Concentration Data

**6.1. Measurements and Errors:** Concentration measurements were made using the same Nu Sapphire MC-ICPMS instruments as were used for the isotopic measurements. Concentration measurements used the most abundant isotope - $^{7}Li$ for lithium, $^{40}Ca$ for calcium etc. All concentration measurements used the initial solution before centrifugation as the reference solution. For example, when measuring the $^{7}Li$ concentration in LiCl at the top and bottom of the centrifuge tube after an experiment, these concentrations used the initial solution of LiCl before centrifugation as the reference. This eliminated errors in the concentration measurement which are due to the presence of different counterions or other impurities.

The combined error in the concentration measurements given is estimated at ±6-7%. This comes from the combined errors of: Sample mass measurement (<1% as 0.1 mg precision used and samples were ~25-75 mg), weighing scales used for water dilution (~2% as 1 mg precision and used down to 50-100 mg), diluting process (<1%), and actual measurement (2-3%). Such error does not affect errors in the isotope ratio, as all errors above cancel out for two isotopes. Therefore, the errors in isotope ratio measurements are still the same as those discussed in section S5.1.

Tables S13 and S14 give the measured concentration results for Ca and Li, respectively.

**Table S13**

| Sample | Time (Hours) | Inner Concentration | Outer Concentration | Isotope Selectivity |
|---|---|---|---|---|
| 0.1 m $CaCl_2$, 1 | **24** | 0.0518 m | 0.160 m | 1.1913 |
| 0.1 m $CaCl_2$, 2 | **24** | 0.0514 m | 0.121 m | 1.2023 |
| 0.1 m $Ca(NO_3)_2$, 1 | **24** | 0.0466 m | 0.191 m | 1.1965 |
| 0.1 m $Ca(NO_3)_2$, 2 | **24** | 0.0462 m | 0.203 m | 1.2011 |
| 0.1 m $CaS_2O_3$, 1 | **24** | 0.0263 m | 0.288 m | 1.2095 |
| 0.1 m $CaS_2O_3$, 2 | **24** | 0.0260 m | 0.260 m | 1.1922 |

**Table S14**

| Sample | Time (Hours) | Inner Concentration | Outer Concentration | Isotope Selectivity |
|---|---|---|---|---|
| 1 m LiCl, 1 | **72** | 0.793 m | - | 1.0447 |
| 1 m LiCl, 2 | **72** | 0.788 m | 1.297 m | 1.0514 |
| 1 m LiBr, 1 | **72** | 0.468 m | 1.812 m | 1.0531 |
| 1 m LiBr, 2 | **72** | 0.455 m | 1.756 m | 1.0531 |
| 1 m LiI, 1 | **72** | 0.273 m | 2.000 m | 1.0544 |
| 1 m LiI, 2 | **72** | 0.279 m | 2.058 m | 1.0526 |
| 1 m $Li_2MoO_4$, 1 (2 m Li) | **72** | 1.144 m | 3.865 m | 1.0396 |
| 1 m $Li_2MoO_4$, 2 | **72** | 1.034 m | - | 1.0410 |
| 2 m LiOH, 1 | **72** | 1.513 m | 2.566 m | 1.0458 |
| 2 m LiOH, 2 | **72** | 1.358 m | 2.496 m | 1.0466 |
| 5 m LiCl, 1 | **72** | 4.818 m | 5.251 m | 1.0458 |
| 5 m LiCl, 2 | **72** | 4.818 m | 5.140 m | 1.0471 |
| 0.1 m LiTFSI, 1 | **24** | 0.0388 m | 0.305 m | - |
| 0.1 m LiTFSI, 2 | **24** | 0.0474 m | 0.295 m | - |
| 0.1 m $Li_2MoO_4$, 1 | **24** | 0.0937 m | 0.408 m | 1.0275 |
| 0.1 m $Li_2MoO_4$, 2 | **24** | 0.0980 m | 0.431 m | 1.0285 |
| 0.1 m $Li_2C_2O_4$ ,1 | **24** | 0.1365 m | 0.301 m | 1.0287 |
| 0.1 m $Li_2C_2O_4$, 2 | **24** | 0.1372 m | 0.279 m | 1.0273 |
| 0.1 m LiOH, 1 | **24** | 0.0844 m | 0.113 m | 1.0210 |
| 0.1 m LiOH, 2 | **24** | 0.0881 m | 0.117 m | 1.0199 |
| 0.5 m $Li_2SO_4$, 1 | **24** | 0.708 m | 1.468 m | 1.0276 |
| 0.5 m $Li_2SO_4$, 2 | **24** | 0.633 m | 1.522 m | 1.0271 |
| 0.5 m $LiNO_3$, 1 | **24** | 0.395 m | 0.614 m | 1.0295 |
| 0.5 m $LiNO_3$, 2 | **24** | 0.387 m | 0.676 m | 1.0304 |
| 1 m LiOH, 1 | **24** | 0.828 m | 1.176 m | - |
| 1 m LiOH, 2 | **24** | 0.875 m | 1.198 m | - |
| 2 m LiOH, 1 | **24** | 1.900 m | 2.625 m | 1.0231 |
| 2 m LiOH, 2 | **24** | 1.944 m | 2.482 m | 1.0236 |
| 5 m LiCl, 1* | **24** | 4.701 m | 4.541 m | 1.0237 |
| 5 m LiCl, 2* | **24** | 4.681 m | 4.740 m | 1.0240 |
| 10 m LiCl, 1* | **24** | 9.755 m | 9.916 m | 1.0159 |
| 10 m LiCl, 2* | **24** | 9.412 m | 9.635 m | 1.0157 |

*It is thought that additional errors were introduce as a result of the 5 m and 10 m solution references partially evaporating and thereby increasing their concentration and artificially lowering the measured sample values. The magnitude of this additional error is estimated to be at most ±5%.

## Section S7. Continuous Process Considerations

### 7.1. Design Considerations:

All experiments performed in this study used a biomedical ultracentrifuge and a swinging-bucket rotor (Beckman Optima XPN-100 Ultracentrifuge and a SW 60 Ti rotor). This configuration is for demonstrating the concept. In practical continuous production, the principles of modern gas centrifuges could be used to create an optimized design. These principles are:

1. An axially symmetric cylindrical centrifuge rotating in a vacuum on a needle bearing and supported by magnetic bearings
2. The flow of solution being continuous in and out of the centrifuge, such that it does not ever deviate from its working rotational speed to accelerate or slow down
3. The use of the countercurrent centrifuge method to multiply the separation factor within each stage
4. A broad cascade of centrifuges, whereby the product of one centrifuge becomes the feed of another in succession, leading to the final overall enrichment

### 7.2. Countercurrent Centrifuge Model:

A countercurrent centrifuge was modeled for isotopes of water (HDO and $H_2{}^{18}O$ inside $H_2{}^{16}O$) according to Eqn. S5.

$$\vec{J}_i = -D_i \nabla c_i + D_i \frac{\omega^2 \vec{\boldsymbol{r}}}{RT} c_i M_i (1 - \bar{v}_i \rho_{soln}) + c_i \vec{\boldsymbol{V}} \quad with \quad \frac{\partial c_i}{\partial t} = -\nabla \cdot (\vec{J}_i) \qquad \text{(S5)}$$

The third term on the right represents advection, where $\vec{\boldsymbol{V}}$ is the velocity field of the flow. For a countercurrent centrifuge this is modeled to be a closed-loop flow contained within a thin stream along the boundaries of the centrifuge. Eqn. S5 was solved numerically in the axial and radial directions with time.

Figure S3 shows the modeling results upon reaching equilibrium. The flow direction was chosen to be counter clockwise so that the heavier isotopes would concentrate at the top of the centrifuge. The modeled system had an inner and outer radius of 5 cm and 10 cm, respectively, and an angular velocity of 50 kRPM. These are both a 16.7% reduction compared to the centrifuge used in this study, and therefore a 30.6% reduction in peripheral speed and 51.8% reduction in wall stress and equilibrium selectivity (under half). Even in this case, the equilibrium separation factor along the axial direction for a 60 cm tall centrifuge is >1.20 per unit mass difference.

Fig. S3a shows the equilibrium distribution of $^{18}O$ throughout the centrifuge. The thin stream of counter-clockwise flow along the boundaries has the effect of concentrating the heavier isotope at the top outer-radii corner. The thin stream travels upwards at the outer radii and downwards at the inner radii. Steep concentration gradients are developed along the axial direction as shown in Fig. S3b, leading to large separation factors in that direction. In the radial direction, typical separations occur as the result of the centrifugal forces (Fig. S3c).

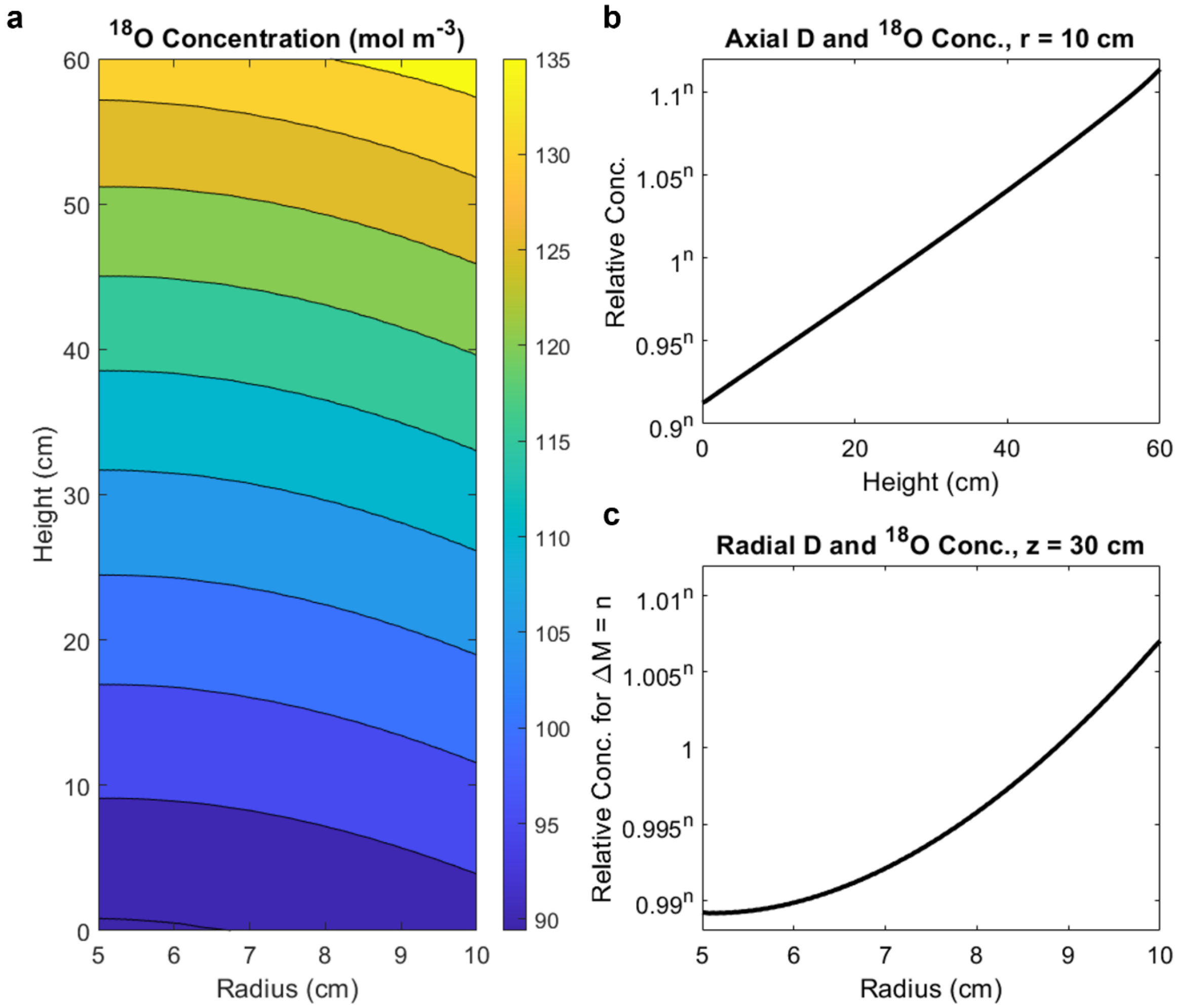

**Figure S3** - The equilibrium countercurrent centrifugation modeling results. (a) The distribution of $H_2{}^{18}O$ throughout the centrifuge. (b) The relative concentration of isotope species as a function of their mass difference from $H_2{}^{16}O$ along the outer radius axial line. (E.g., $\Delta M = 1$ from HDO). (c) The relative concentration of isotope species along the z = 30 cm horizontal centerline.

## Section S8. Throughput Analysis

The separative throughput of a centrifuge is proportional to the product of the concentration and diffusion coefficient of the target species, i.e., J ∝ c×D (*11*). The number of cascade stages in a centrifuge is inversely proportional to ln(α) ∝ ΔM. So J × ln(α) is also a meaningful parameter to evaluate the separation power of a system.

### 8.1. Comparison of Gas and Liquid Centrifuge:

For gases, the diffusivity can be modeled using Chapman–Enskog theory as in Eqn. S6

$$D = \frac{AT^{3/2}}{p\sigma_{12}^2 \Omega}\sqrt{\frac{1}{M}} \qquad \text{(S6)}$$

where $A$ is an empirical coefficient, $T$ is the absolute temperature, $M$ is the molar mass, $p$ is the pressure, $\sigma_{12}$ is the average collision diameter and $\Omega$ is a temperature-dependent collision integral. As $p = cRT$ based on the ideal gas law, c×D is a constant at a given temperature so that changes in the pressure and gas concentration do not affect this parameter (*32*). Moreover, the temperature dependence of c×D is $T^{1/2}$, which is a weak dependence. From 25 to 100 °C, c×D only increases by ~12% for a gas, as given in Figure S4a. This originates from the fact that *c* decreases at a higher temperature and the same pressure.

In contrast, for species in a liquid solution, the product c×D is itself a non-linear function of the solute concentration and can therefore be maximized for a given solute. c×D also increases more significantly in an aqueous solution than gas. As shown in Figure S4a, between 25 and 100 °C, the diffusion coefficient increases ~ 2-2.5% $K^{-1}$ of that at 25°C, which is much stronger than gas (*25*). The solubility can also increase by 50 – 200 % compared to that at 25 °C. Hence, c×D can increase by 50-100% from 25 to 50 °C, and even 2-5x from 25 to 100 °C. The exact increase is solution-specific and can be optimized by tuning the chemical composition of solutes and solvents.

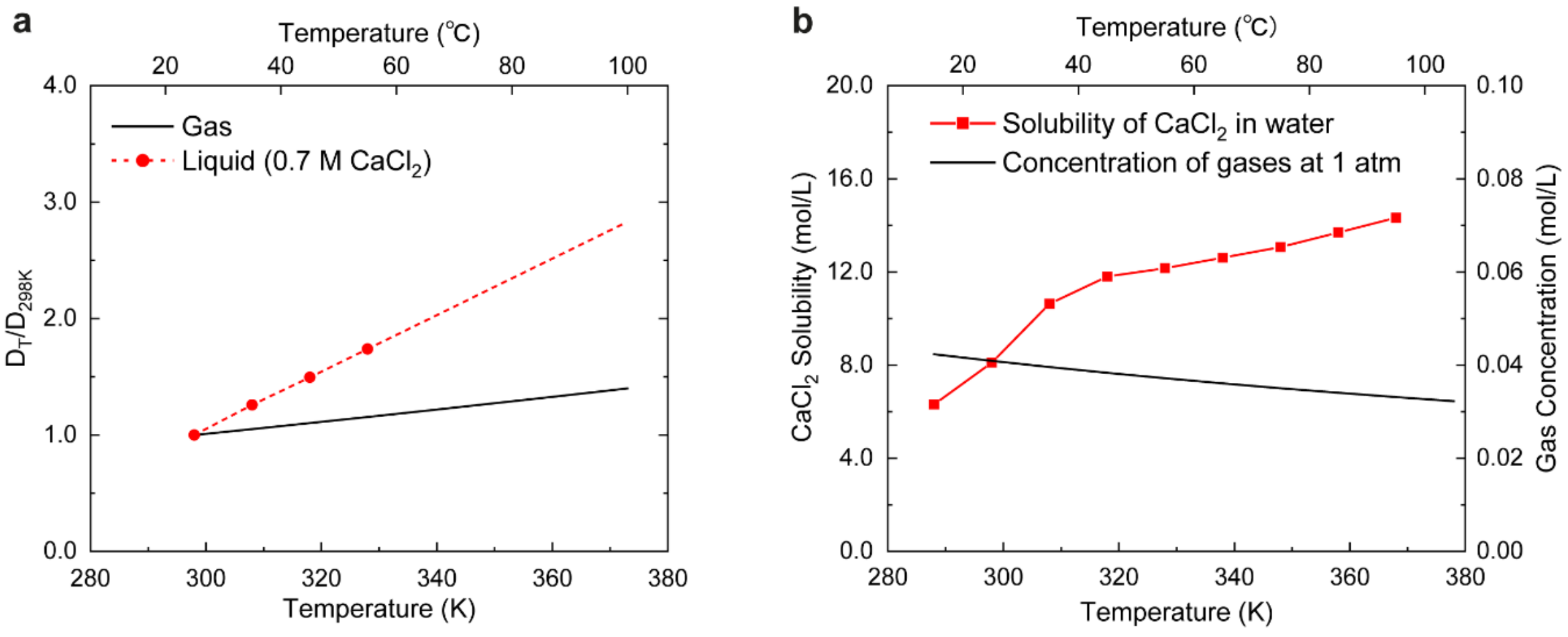

**Figure S4** – (a) The temperature dependence of diffusivity for gas and liquid (0.7 M $CaCl_2$ aqueous solution). (b) The temperature dependence of $CaCl_2$ solubility in water and the temperature dependence of concentration for gaseous species at 1 atm.

### 8.2. Summary of throughput values for gas and liquid centrifuge:

Table S16 gives some values for c×D and c×D×ln(α) at two concentrations and temperatures for the some of the salts used in this study, and the flux comparison between gaseous and liquid species is provided. The term ln(α) is represented by ΔM/T, as the selectivity is an exponential function of the isotope mass difference and the inverse temperature. At steady state, the flux of liquid centrifugation is about ~1/10 of gas centrifugation at moderate temperature but this could be elevated with higher temperatures. Even at 50°C the separation for $^{40}Ca/^{48}Ca$ and $^{35}Cl/^{37}Cl$ show good throughput potential, and especially $^{1}H/^{2}H$ and $^{16}O/^{18}O$ due to the high solvent concentration. Most importantly, all 4 of these isotope pairs could be simultaneously separated using $CaCl_{2(aq)}$.

**Table S15**

| Gas | Temperature (°C) | c×D×10⁶ (∝ Flux) (mol m⁻¹ s⁻¹) | c×D×ln(α)×10⁹ (mol m⁻¹ s⁻¹) |
|---|---|---|---|
| $UF_6$ | 25 | 61.7 | **621** |
| | 50 | 66.5 | **617** |
| | 75 | 71.2 | **614** |

**Table S16**

| Liquid Solution | Temperature (°C) | Concentration (mol L⁻¹) | Diffusion Coefficient ($10^9$ m² s⁻¹) | $c_i$×$D_i$ ×10⁶ (∝ Flux) (mol m⁻¹ s⁻¹) | $c_i$×$D_i$ ×ln(α)×10⁹ (mol m⁻¹ s⁻¹) |
|---|---|---|---|---|---|
| **Ca**$Cl_{2(aq)}$ | 25 | 1.0 M | 1.22 | 1.22 | 32.7 |
| | | 5.0 M | 0.72 | 3.60 | 96.6 |
| | 50 | 1.0 M | 2.06 | 2.06 | 51.0 |
| | | 5.0 M | 1.21 | 6.05 | 150 |
| | 75 | 1.0 M | 3.48 | 3.48 | 80.0 |
| | | 5.0 M | 2.07 | 10.35 | 237.9 |
| | 100 | 1.0 M | 5.87 | 5.87 | 126.0 |
| | | 5.0 M | 3.49 | 17.45 | **374** |
| **Li**$Cl_{(aq)}$ | 25 | 1.0 M | 0.93 | 0.93 | 3.1 |
| | | 5.0 M | 0.62 | 3.44 | 11.5 |
| | 50 | 1.0 M | 1.50 | 1.50 | 4.6 |
| | | 5.0 M | 0.91 | 5.10 | **15.8** |
| Li**Cl**$_{(aq)}$ | 25 | 1.0 M | 1.68 | 1.68 | 11.3 |
| | | 5.0 M | 1.05 | 5.25 | 35.2 |
| | 50 | 1.0 M | 2.47 | 2.47 | 15.3 |
| | | 5.0 M | 1.54 | 7.70 | **47.7** |
| **$H_2$**O | 25 | 110.8 M | 2.32 | 257 | 863 |
| | 50 | 109.8 M | 3.89 | 427 | 1322 |
| | 75 | 108.4 M | 5.93 | 643 | **1847** |
| $H_2$**O** | 25 | 55.4 M | 2.32 | 129 | 863 |
| | 50 | 54.9 M | 3.89 | 214 | 1322 |
| | 75 | 54.2 M | 5.93 | 321 | **1847** |
| **$K_2$**$MoO_{4(aq)}$ | 25 | $c_{K+}$=1.0 M | 1.891 | 1.89 | 12.7 |
| | | $c_{K+}$=5.0 M | 1.249 | 6.99 | 46.9 |
| | 50 | 1.0 M | 3.039 | 3.04 | 18.8 |
| | | 5.6 M | 1.85 | 10.40 | **64.4** |
| $K_2$**Mo**$O_{4(aq)}$ | 25 | 1.0 M | 0.694 | 0.69 | 18.5 |
| | | 5.0 M | 0.407 | 2.04 | 54.7 |
| | 50 | 1.0 M | 1.172 | 1.17 | 29.0 |
| | | 5.0 M | 0.688 | 3.44 | **85.2** |

## Section S9. Cost Analysis

A preliminary cost analysis can be made for liquid solution centrifugation based on its similarities to the gas centrifuge method.

Available data from the Energy Information Administration (EIA) 2021 Uranium Marketing Annual Report indicates the average price of separative work unit (SWU) for uranium isotopes in a gas centrifuge is $100 from 2020-2021 (*26*). The reported cost estimates for gas centrifuge plants indicate that the power consumption is about 62 kWh/SWU, corresponding to ~$7/SWU (Table S17) (*27*). The remaining $93/SWU is attributed to the capital and operational costs which are assumed as the same for both gas and liquid centrifugation methods. Under this assumption, the main cost difference between the two methods arises from the power requirements of operation, which can be attributed to the frictional losses of the bearing and scoops, as well as keeping the vacuum (*11*). Therefore, the power costs will be proportional to the centrifugation time, and so analogous power requirements can be made. Taking calcium as an example (e.g., 5 M $CaCl_2$ water at 50°C), the centrifugation time is inversely proportional to the product of the flux of the raw materials. A direct comparison can then be made with $UF_{6(g)}$ from Table 15.

$$t_{\text{multiplier}} = \left(\frac{J_{CaCl_2}}{J_{UF_6}}\right)^{-1} = \frac{J_{UF_6}}{J_{CaCl_2}} = \frac{6.53 \times 10^{-5}\, mol^{-1} \cdot m \cdot s}{6.05 \times 10^{-6}\, mol^{-1} \cdot m \cdot s} = 10.8$$

Therefore, the centrifugation time for $^{40}Ca/^{48}Ca$ is about 10.8 times longer than for uranium for the same molar throughput, with a corresponding power cost of 10.8 times larger per mole. Since fluxes are typically expressed in molar quantities but separative work units are per kg, it is necessary to convert between them using the molar mass of the element. The unit of Molar Separative Work (MSW) is then introduced for comparison, as moles are typically used in scientific contexts. 1 MSW is defined as 1 mole of separative work, and therefore the 1 MSW = 0.238 SWU for Uranium.

The gas centrifuge costs for uranium isotopes are then $23.8 per MSW (0.238 kg/mol × $100), with $1.67 being the power costs and $22.13 being the capital/operational costs. The costs for $^{40}Ca/^{48}Ca$ separation are then (10.8×$1.67 + $22.13) = $40.17 per mole of separative work (MSW).

Larger separation factors $\alpha$ per stage will increase the overall production output proportionally. This will benefit isotopes pairs with a large $\Delta M$ and disadvantage those with $\Delta M = 1$ or 2 Da comparatively. As shown in Section S4, $\alpha$ is exponential with $\Delta M$, and the number of cascade stages is inversely proportional to $\ln(\alpha)$. Hence, the separation of $^{40}Ca/^{48}Ca$ is 8/3 = 2.67 times as efficient as for $^{235}U/^{238}U$ for a given centrifuge cascade, resulting in a cost of $40.2 / 2.67 = $15.1 / MSW output.

Similar cost estimations for enriching isotopes using liquid centrifugation can be applied to other elements. The analysis does not account for the possibility of using the same salt to separate the isotopes of multiple elements, as well as those in the solvent, which would improve overall cost effectiveness.

It is acknowledged that the assumptions made in this analysis are broad and that real production costs will vary greatly, particularly as centrifuge designs will differ and that handling liquids and gases is not analogous. It is hoped that this acts simply as a first order-of-magnitude approximation for the potential costs of such a process.

**Table S17** – Levelized SWU costs, operating centrifuge capacity (Europe and Japan) (5% cost of capital, 6.51% capital recovery factor, +0% IDC, 0% contingency). Reproduced with permission from (*27*) - G. Rothwell, Market Power in Uranium Enrichment. *Sci. Glob. Secur.* **17**, 132-154 (2009).

| Firm<br>Plant | (2008$) | Urenco<br>Capenhurst | Urenco<br>Almelo | Urenco<br>Gronau | JNFL<br>Rokkasho |
|---|---|---|---|---|---|
| **Plant capacity** | t SWU/yr | **3,400** | **2,900** | **1,800** | **1,500** |
| Overnight cost | $M | $2,342 | $2,076 | $1,445 | $1,095 |
| Total capital invest cost | $M | $2,342 | $2,076 | $1,445 | $1,095 |
| **Capital/SWU** | $/SWU | **$44.82** | **$46.56** | **$52.21** | **$56.98** |
| Staff size | people | 340 | 317 | 257 | 219 |
| Annual fully burden salary | $k/yr | $120 | $120 | $120 | $120 |
| **Labor/SWU** | $/SWU | **$11.99** | **$13.10** | **$17.12** | **$20.99** |
| Electricity consumption | kWh/SWU | 62 | 62 | 62 | 62 |
| Electricity price | $/MWh | $107 | $107 | $107 | $107 |
| **Electricity/SWU** | $/SWU | **$6.65** | **$6.65** | **$6.65** | **$6.65** |
| **Materials/SWU** | $/SWU | **$6.89** | **$7.16** | **$8.03** | **$8.76** |
| **Annual total costs** | $M | **$239** | **$213** | **$151** | **$117** |
| **Levelized SWU cost** | $/SWU | **$70** | **$73** | **$84** | **$93** |

## Section S10: Solution Non-idealities

Debye-Hückel theory treats the solvent as only a mediator for the electrostatic interactions between the dissolved ions in a solution. Therefore, the only property of importance is the dielectric constant, $\epsilon_r$, as this relates to the electric field strength away from a charge, as given in Eqn. S7, where all terms are defined in (*29*). Since this is a constant for a given solvent, it can be factored out. Therefore, the deviation of the thermodynamic factor away from one is inversely proportional to the dielectric constant, as in Eqn. S8.

$$ln\left(\gamma_i^{DH}\right) = \frac{-q_i^2}{8\pi\epsilon_r\epsilon_o k_B T(R_i + l_D)} \qquad (S7)$$

$$\vartheta = 1 + c\frac{\partial ln(\gamma)}{\partial c} \quad \rightarrow \quad \vartheta - 1 = \frac{1}{\epsilon_o} c \frac{\partial}{\partial c}\left[\frac{-q_i^2}{8\pi\epsilon_r k_B T(R_i + l_D)}\right] \qquad (S8)$$

It must be noted that Debye-Hückel theory and its extensions are only valid to low concentrations of <0.1 M or <0.5 M where the association of ions is not significant. Beyond this, ion association and other neglected factors become large and the predictions have not agreed with experiments (*29*). No theory yet proposed has been able to accurately quantify activity coefficients for aqueous solutions at high salt concentration, and almost no attention has been given to organic solvents.

## References and Notes:

1. Meeting Isotope Needs and Capturing Opportunities for the Future: The 2015 Long Range Plan for the DOE-NP Isotope Progarm, *NSAC Isotopes Subcommitee, July 2015* (2015).
2. J. W. Beams, L. B. Snoddy, A. R. Kuhlthau, Tests of the theory of isotope separation by centrifuging. 2$^{nd}$ U.N. Conference on the Peaceful Uses of Atomic Energy. **4**, 428-434 (1958).
3. L. O. Love, Electromagnetic Separation of Isotopes at Oak Ridge. *Science.* **182**, 10 (1973).
4. T. Graham, On the molecular mobility of gases. *Phil. Trans. R. Soc* **153**, 385-405 (1863).
5. H. C. Urey, The thermodynamic properties of isotopic substances. *J. Chem. Soc.* **1**, 562-581 (1947).
6. V. S. Letokhov, Laser isotope separation. *Nature.* **277**, 605-610 (1979).
7. R. L. Murray, K. E. Holbert, *Nuclear Energy (Eighth Edition), Chapter 15 - Isotope Separators*. (Wiley, 2020).
8. A. N. Cheltsov, L. Y. Sosnin, V. K. Khamylov, Centrifugal enrichment of nickel isotopes and their application to the development of new technologies. *J Radioanal Nucl Chem.* **299**, 981–987 (2014).
9. J. Bigeleisen, M. G. Mayer, Calculation of Equilibrium Constants for Isotopic Exchange Reactions. *J. Chem. Phys* **15**, (1947).
10. L. Onsager, R. M. Fuoss, Irreversible processes in electrolytes. diffusion, conductance and viscous flow in arbitrary mixtures of strong electrolytes. *J. Phys. Chem.* **36**, 2689-2778 (1932).
11. H. W. Hsu, [Chapter III] in *Separations by centrifugal phenomena*. (Wiley, 1981). pp. 50-54.
12. T. Osawa, M. Ono, F. Esaka, S. Okayasu, Y. Iguchi, T. Hao, M. Magara, T. Mashimo. Mass-dependent isotopic fractionation of a solid tin under a strong gravitational field. *EPL.* **85**, 6, 64001 (2009).
13. T. Mashimo, M. Ono, X. Huang, Y. Iguchi, S. Okayasu, K. Kobayashi, E. Nakamura, Isotope separation by condensed matter centrifugation: Sedimentation of isotope atoms in Se. *J. Nucl. Sci. Technol.* **45**, 6, 105-107 (2014).
14. M. Ono, T. Mashimo, Sedimentation process for atoms in a Bi-Sb system alloy under a strong gravitational field: A new type of diffusion of substitutional solutes. *Philos. Mag A.* **82**, 3, 591-600 (2002).
15. X. Ge, X. Wang, M. Zhang, S. Seetharaman, Correlation and prediction of activity and Osmotic coefficients of aqueous electrolytes at 298.15 K by the modified TCPC model. *J. Chem. Eng. Data.* **52**, 2, 538-547 (2007).
16. N. Xin, Y. Sun, C. J. Radke, J. M. Prausnitz, Osmotic and activity coefficients for five lithium salts in three non-aqueous solvents. *J. Chem. Thermodyn.* **132**, 83-92 (2019).
17. D. Brugge, J. L. deLemos, C. Bui, The Sequoyah Corporation fuels release and the church rock spill: Unpublicized nuclear releases in American Indian Communities. *Am. J. Public Health.* **97**, 9, 1595-1600 (2007).
18. B. L. Zaret, F. J. Wackers, Nuclear Cardiology. *N Engl J Med.* **329**, (1993).
19. A. A. Palko, J. S. Drury, G. M. Begun, Lithium isotope separation factors of some two-phase equilibrium systems. *J. Chem. Phys.* **64**, 1828-1837 (1976).
20. D. Zucker, J. S. Drury, Separation of calcium isotopes in an Amalgam system. *J. Chem. Phys.* **41**, 1678-1681 (1964).
21. A. Rittirong, T. Yoshimoto, R. Hazama1, T. Kishimoto, T. Fujii, Y. Sakuma, S. Fukutani, Y. Shibahara, A. Sunaga, Isotope separation by DC18C6 crown-ether for neutrinoless double beta decay of $^{48}Ca$. *J. Phys.: Conf. Ser.* **2147**, (2022).
22. W. M. Rutherford, K. W. Laughlin, Separation of Calcium Isotopes by Liquid Phase Thermal Diffusion. *Science.* **211**, 1054-1056 (1981).
23. W. Dai, F. Moynier, M. Paquet, J. Moureau, B. Debret, J. Siebert, Y. Gerard, Y. Zhao, Calcium isotope measurements using a collision cell (CC)-MC-ICP-MS. *Chem. Geol.* **590**, (2022).
24. H. Chen, N. J. Saunders, M. Jerram, A. N. Halliday, High-precision potassium isotopic measurements by collision cell equipped MC-ICPMS. *Chem. Geol.* **578**, (2021).

25. E. A. Hollingshead, A. R. Gordon, The Differential Diffusion Constant of Calcium Chloride in Aqueous Solution. *J. Chem. Phys.* **9**, (1941).
26. Uranium Marketing Annual Report United States Energy Information Administration (EIA), (2022).
27. G. Rothwell, Market Power in Uranium Enrichment. *Sci. Glob. Secur.* **17**, 132-154 (2009).
28. M. Blau, R. Ganatra, M. A. Bender, 18F-fluoride for bone imaging. *Seminars in Nuclear Medicine.* **2**, (1972).
29. J. L. Liu, C. L. Li, A generalized Debye-Hückel theory of electrolyte solutions. *AIP Adv.* **9**, (2019).
30. F. A. Lindemann, F. W. Aston, The possibility of separating isotopes. *Phil. Mag.* **37**, 523-534 (1919).
31. G. J. Hooyman, *"Thermodynamics of Diffusion and Sedimentation" in Ultracentrifugal Analysis in Theory and Experiment*. (J. W. Williams, Ed. (Academic Press, 1963), 1962), pp. 3-12.
32. M. Brown, E. G. Murphy, Measurements of the self-diffusion coefficient of uranium hexafluoride. *Trans. Faraday Soc.* **61**, 2442-2446 (1965).